\documentclass[12pt, draftclsnofoot, onecolumn]{IEEEtran}
\usepackage{graphicx}
\usepackage{amsmath}
\usepackage{amssymb}
\usepackage[compress]{cite}
\usepackage{color}
\usepackage{subfigure}

\newtheorem{prop}{Proposition}
\newtheorem{rem}{Remark}
\newcommand{\figsize}{0.37}

\newcommand{\spp}{\textrm{sp}}

\newcommand{\mar}{\textcolor{black}}
\newcommand{\tcb}{\textcolor{black}}

\begin{document}

\title{Link Selection in Hybrid RF/VLC Systems\\ under Statistical Queueing Constraints}
\author{Marwan Hammouda, Sami Ak{\i}n$^\dagger$, Anna Maria Vegni, Harald Haas, and J\"{u}rgen Peissig

\thanks{M. Hammouda, S. Ak{\i}n, and J. Peissig are with the Institute of Communications Technology, Leibniz Universit\"{a}t Hannover, 30167 Hanover, Germany. E-mails: \{marwan.hammouda, sami.akin, and peissig\}@ikt.uni-hannover.de.}
\thanks{A. M. Vegni is with COMLAB laboratory, Department of Engineering, Roma TRE University, 00146 Roma, Italy. E-mail: annamaria.vegni@uniroma3.it.}
\thanks{H. Haas is with the Institute for Digital Communications, Li-Fi Research and Development Centre, the University 
of Edinburgh, Edinburgh, UK, EH9 3JL. E-mail: h.haas@ed.ac.uk.}
\thanks{$\dagger$ This work was supported by the European Research Council under Starting Grant--306644.}
}

\maketitle
\vspace{-1.0cm}
\begin{abstract}
The co-deployment of radio frequency (RF) and visible light communications (VLC) technologies has been investigated in indoor environments to enhance network performances and to address specific quality-of-service (QoS) constraints. In this paper, we explore the benefits of employing both technologies when the QoS requirements are imposed as limits on the buffer overflow and delay violation probabilities, which are important metrics in designing low latency wireless networks. Particularly, we consider a multi-mechanism scenario that utilizes RF and VLC links for data transmission in an indoor environment, and then propose a link selection process through which the transmitter sends data over the link that sustains the desired QoS guarantees the most. Considering an ON-OFF data source, we employ the maximum average data arrival rate at the transmitter buffer and the non-asymptotic bounds on data buffering delay as the main performance measures. We formulate the performance measures under the assumption that both links are subject to average and peak power constraints. Furthermore, we investigate the performance levels when either one of the two links is used for data transmission, or when both are used simultaneously. Finally, we show the impacts of different physical layer parameters on the system performance through numerical analysis. 
\end{abstract}

\section{Introduction}
The ever-growing demand for mobile communications triggered a quest for technical solutions that will support stringent quality-of-service (QoS) constraints. Thanks to the significant advances in white light emitting diodes (LEDs) research, and the availability of an extensive unregulated spectrum, visible light communication (VLC) has emerged as a promising technology. We can utilize LEDs simultaneously for data transmission and illumination, since they have many unique aspects compared to the other communication technologies~\cite{rahaim2011hybrid}. \mar{Moreover, we can improve data security because light does not penetrate the surrounding walls.} We can also sustain an all-important \emph{green} agenda and minimize the installation costs because we do not require an extensive infrastructure. Nevertheless, attention must be paid to certain limitations and challenges in VLC systems, e.g., smaller coverage, strong dependence on line-of-sight components and achievable rates that vary with spatial fluctuations~\cite{basnayaka2017design}. In order to overcome these constraints, researchers proposed hybrid RF/VLC 
systems~\cite{rahaim2011hybrid, chowdhury2014cooperative, basnayaka2015hybrid, bao2014protocol, shao2014indoor, kashef2016energy, vegni2012handover, wang2015efficient, liang2015novel, shao2015delay, li2015cooperative,wang2015dynamic}, where end users can benefit from the wide coverage area that RF systems support and the stable rates that VLC systems provide. Such networks are practically feasible as RF and VLC systems can coexist without causing interference on each other and operate in the same environment, such as offices and rooms.

Comparing hybrid RF/VLC systems with systems that employ either RF or VLC only, the authors in~\cite{chowdhury2014cooperative, basnayaka2015hybrid, stefan2013area, kashef2016energy, rahaim2011hybrid} demonstrated a remarkable increase in data transmission throughput, energy efficiency and delay performance in hybrid RF/VLC systems. Moreover, the authors in~\cite{bao2014protocol, shao2014indoor} projected a hybrid system in which they use VLC links for down-link communication and RF links for up-link communication. In such a system, the authors in~\cite{vegni2012handover, wang2015efficient, liang2015novel} and the ones in~\cite{li2015cooperative,wang2015dynamic,wang2017optimization} investigated handover and load balancing mechanisms, respectively. Alternatively, considering an outdoor environment, the authors in~\cite{kazemi2014outage} studied a point-to-point transmission scenario in which the system can switch between RF links and VLC links after comparing the signal-to-noise ratio levels in each link. Regarding the same system setting, the authors in~\cite{chatzidiamantis2011diversity} assumed that both RF and VLC links have the same transmission rates, and then proposed a diversity-based transmission scheme such that the transmitter sends data by employing both links simultaneously.

The aforementioned studies analyzed the hybrid RF/VLC systems mostly from the physical layer perspective, i.e., they did not concentrate on the data link layer metrics, such as limits on the buffer overflow and buffering delay probabilities, as much as needed. Noting the dramatic increase in the demand for reliable delay-sensitive services in recent years (mobile video traffic making up $55\%$ of the total global data traffic at the end of $2014$~\cite{Cisco}), in addition to the physical layer performance metrics, we need QoS metrics that can be a cross-layer analysis tool between the physical layer features and the performance levels in data-link layer. In this context, the authors in~\cite{chang1994stability, wu2003effective, choudhury1996squeezing,kelly1996notes, akin2010effective, hammouda2014effective, ozmen2016wireless} performed cross-layer analysis between physical and data-link layers in many different RF scenarios. Regarding a Markovian data arrival process at a transmitter buffer and a statistically varying data service process in a wireless channel, the authors in~\cite{ozmen2016wireless} characterized the maximum average data arrival rate at the transmitter buffer with the presence of statistical constraints on the buffer overflow probability. On the other hand, to the best of our knowledge, there are only a few studies that investigated cross-layer performance levels in VLC systems. For example, the authors employed effective capacity as a performance measure in resource allocation schemes in VLC systems~\cite{jin2016resource} and heterogeneous networks composed of VLC and RF links~\cite{jin2015resource}. Here, we note that the authors in~\cite{jin2015resource} concentrated on the case of constant data arrival rates at the transmitter buffer, which is not realistic in certain practical settings. For more details in effective capacity, we refer to~\cite{wu2003effective}.

In this paper, assuming an ON-OFF modeled data arrival process at the transmitter buffer and that the transmitter can use both RF and VLC channels for data transmission, we investigate the performance gains achieved by a hybrid RF/VLC system. This operates under statistical QoS constraints, which are inflicted as limits on the buffer overflow and delay violation probabilities. We perform a cross-layer analysis of these hybrid systems in physical and data-link layers. We employ first the maximum average data arrival rate at the transmitter buffer considering the asymptotic buffer overflow probability approximation, and then non-asymptotic buffering delay violation probability as the main performance measures. We propose a mathematical toolbox to system designers for performance analysis in hybrid RF/VLC systems that work under low latency conditions. To summarize, the main contributions of this paper are as follows:

\begin{enumerate}
	\item Assuming that both RF and VLC links are subject to average and peak power constraints, we express the maximum average data arrival rate at the transmitter that the data service process from the transmitter to the receiver can support under QoS constraints when either the RF or VLC link is used, or both links are simultaneously used for data transmission.
	\item We propose three different link usage strategies. We base two of the proposed strategies on the assumption that the receiver does not have a multihoming capability, thus data transmission is possible over only one link, either the RF or VLC link. In the third strategy, we assume that link aggregation is possible and data can be transmitted over both links simultaneously following a power sharing policy. 
	\item We obtain the non-asymptotic data backlog and buffering delay violation probability bounds considering the proposed link usage strategies.
\end{enumerate}

\mar{Particularly, we provide a rudimentary model for multi-mechanisms in communication systems that operate under QoS constraints. We employ RF and VLC links as two different mechanisms. Our model can be easily invoked in settings with more than two different mechanisms as well. The reason behind multi-mechanisms in communications is to boost performance levels through the increased degree of freedom. Therefore, in order to introduce our model smoothly and make it easier for readers to understand our objective, we also benefit from the existing literature in RF and VLC studies. However, to the best of our knowledge, the analytical framework provided in this paper, in which we investigate the QoS performance, is not addressed in other studies. One aspect of this hybrid system is that VLC links provide time-invariant transmission rates, while RF links provide rates that vary over time. A communication setting that depends solely on an RF link may suffer low transmission rates, and have longer data backlogs in the transmitter buffer. However, a communication setting that can utilize both RF and VLC links, for instance, can take advantage of the constant transmission rate in the VLC link when the transmission rate in the RF link falls below a certain level.} The rest of the paper is organized as follows. We introduce the hybrid RF/VLC system in Section \ref{sec:System_Model}. We provide detailed descriptions of both RF and VLC channels. In Section \ref{sec:Performance_Analysis}, we provide the performance analysis and the link selection process. In Section \ref{sec:sim}, we substantiate our results with numerical demonstrations. We conclude the paper in Section \ref{sec:conc} and relegate the proofs to the Appendix.

\section{System Model}\label{sec:System_Model}
\begin{figure}
	\centering
	\includegraphics[width=0.3\textwidth]{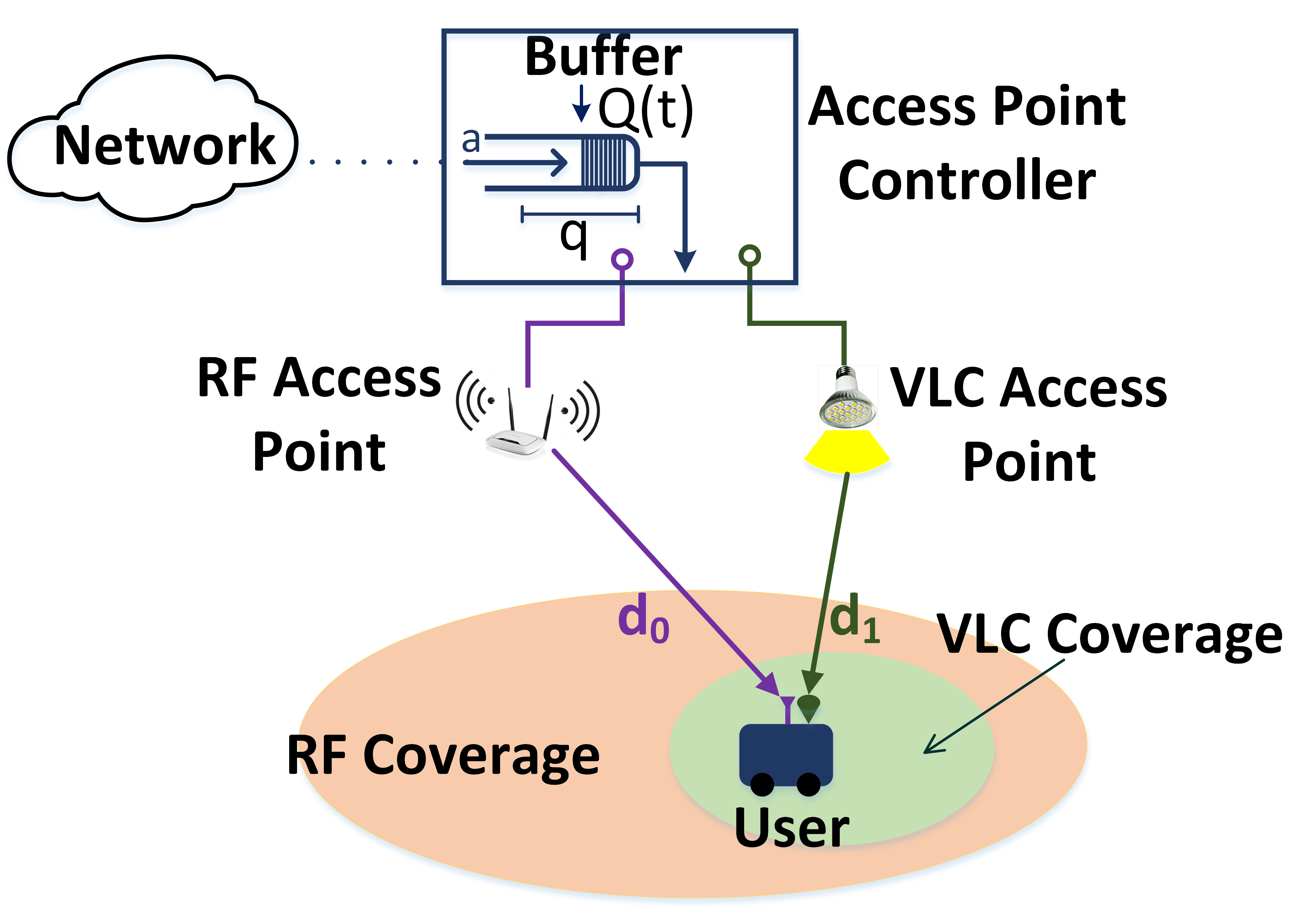}
	\caption{Hybrid RF/VLC system.}\label{Channel_Model}
	\vspace{-0.8cm}
\end{figure}

We consider a network access controller that provides a connection to a user\mar{\footnote{\mar{The analytical framework provided in this paper can easily be extended to a multi-user scenario. For more details, we refer to Remark \ref{rem:multi} in Section III.}}}, through either an RF access point or a VLC access point, which are positioned at different locations in an indoor environment as seen in Figure \ref{Channel_Model}. Herein, we assume a down-link scenario, i.e., the network access controller acts as a transmitter and the user acts as a receiver. In the sequel, we use \textit{transmitter} and \textit{receiver} instead of \textit{network access controller} and \textit{user}, respectively. Initially, the transmitter receives data from a source (or sources) and stores it as packets in its buffer. Subsequently, it sends the data packets in frames of $T$ seconds to the receiver following a given transmission strategy. The receiver is considered to be equipped with an RF front-end and a photo-diode. We also note that the VLC coverage area is generally smaller than the RF coverage area, and that they overlap\mar{\footnote{\mar{This model is also considered in~\cite{wang2015efficient}.}}}. Finally, we consider a power-limited system and assume that the network controller is constrained by a fixed average power budget, denoted as $P_{\text{avg}}$, for data transmission. In the sequel, we initially introduce the RF and VLC channels, and then describe the data source model.

%\vspace{-0.3cm}
\subsection{RF Channel Model}\label{sec:RF_Channel_Model}
During the data transmission in the flat-fading RF channel, the input-output relation at time instant $t$ is expressed as
\begin{equation}\label{input_output_RF}
	y_{r}(t)=x_{r}(t)h(t)+w_{r}(t), 
\end{equation}
where $x_r(t)$ and $y_r(t)$ are the complex channel input and output at the RF access point of the transmitter and the RF front-end of the receiver, respectively. The complex channel input is subject to an average power constraint, $P_{\text{avg},r}$, i.e., $\mathbb{E}\{|x_r(t)|^2\}\leq P_{\text{avg},r}$, and a peak power constraint, $P_{\text{peak},r}$, i.e., $|x_r(t)|^2\leq P_{\text{peak},r}$. Above, $h(t)$ is the complex channel fading gain with an arbitrary distribution having a finite average power, i.e., $\mathbb{E}\{|h(t)|^2\}<\infty$. Furthermore, we consider a block-fading channel and assume that the fading gain stays constant during one transmission frame ($T$ seconds), i.e., \mar{$h(lT)=h(lT+\frac{1}{B_r})=\cdots=h((l+1)T-\frac{1}{B_r})=h_{l}$}, where the available bandwidth is $B_r$ Hz in the channel and $h_{l}$ is the channel fading gain in the $l^{\text{th}}$ time frame. Note that there are $TB_r$ symbols transmitted in one time frame. Moreover, the fading gain changes independently from one frame to another. Meanwhile, $w_r(t)$ is the additive noise at the RF front-end of the receiver, which is a zero-mean, circularly symmetric complex Gaussian random variable with variance $\sigma_{r}^2$, i.e., $\mathbb{E}\{|w_{r}(t)|^2\}=\sigma_{r}^2<\infty$. The noise samples $\{w_{r}(t)\}$ are assumed to be independent and identically distributed.

We assume that a reliable data transmission exists as long as the transmission rate in the channel is lower than or equal to the instantaneous mutual information between the channel input and output\mar{\footnote{\mar{This assumption is based on the known result in the literature that transmitting data at rates less than or equal to the instantaneous mutual information has a high reliability, thus the decoding error is negligible~\cite[Eq. (10)]{letzepis2009outage} \cite{ozarow1994information}.}}}. In particular, when the transmission rate in the $l^{\text{th}}$ time frame (i.e., $R_{l}$ bits per frame) is lower than or equal to the instantaneous mutual information (i.e., $C_{r,l}$ bits per frame) between the channel input \mar{$[x_{r}(lT),x_{r}(lT+\frac{1}{B_r}),\cdots,x_{r}((l+1)T-\frac{1}{B_r})]$} and the channel output \mar{$[y_{r}(lT),y_{r}(lT+\frac{1}{B_r}),\cdots,y_{r}((l+1)T-\frac{1}{B_r})]$}, a reliable data transmission occurs and $R_{l}$ bits are decoded correctly by the receiver. Here, we assume that $TB_r$ is large enough so that the decoding error probability is negligible when $R_{l}\leq C_{r,l}$. In~\cite{shamai1995capacity}, a lower bound on the maximum mutual information (or channel capacity) is provided, where the input has a two-dimensional circularly truncated Gaussian distribution. Therefore, we set the instantaneous data transmission rate to the lower bound and assume that the input has a two-dimensional circularly truncated Gaussian distribution. Specifically, we have a reliable transmission in the $l^{\text{th}}$ time frame when
\begin{subequations}\label{rate_set_lower_bound}
	\begin{equation}
		R_{l}=TB_r\log_{2}\left\{1+\frac{2|h_{l}|^2}{a\sigma_{r}^2}\exp\left\{\frac{bP_{\text{avg},r}}{2}-1\right\}\right\}\leq C_{r,l}\quad\textrm{bits per frame}, \tag{\ref{rate_set_lower_bound}}
	\end{equation}
	where $a$ and $b$ are the solutions of the following equations:
	\begin{align}
		\frac{a}{b}\left[1-\exp\left\{-\frac{bP_{\text{peak},r}}{2}\right\}\right]&=1,\label{eq:a_a}
		\intertext{and}
		2\frac{a}{b}(bP_{\text{peak},r})^{-1}\left[1-\exp\left\{-\frac{bP_{\text{peak},r}}{2}\right\}\left(1+\frac{bP_{\text{peak},r}}{2}\right)\right]&=\frac{P_{\text{avg},r}}{P_{\text{peak},r}}.\label{eq:a_b}
	\end{align}
\end{subequations}
Above, $C_{r,l}$ is time-dependent and changes from one time frame to another because the maximum instantaneous mutual information in each frame is a function of the channel fading gain.
%\vspace{-0.3cm}
\subsection{VLC Channel Model}\label{sec:VLC_Channel_Model}
We assume that the transmitter employs \textit{intensity modulation/direct detection}. Principally, the VLC access point of the transmitter is equipped with an LED and the data is modulated on the intensity of the emitted light. The receiver that collects light using a photo-diode generates an electrical current or voltage proportional to the intensity of the received light. Besides, we know that VLC channels are typically composed of line-of-sight as well as multi-path components. However, the majority of the collected energy at photo-diodes (more than $95$\%) comes from the line-of-sight components in typical indoor scenarios~\cite{komine2004fundamental}. Therefore, we can assume that the VLC channel is flat with a dominant line-of-sight component~\cite{kahn1997wireless, pohl2000channel, 7390429}, and the channel gain does not vary during the data transmission as long as the receiver is stationary\mar{\footnote{\mar{Small-scale variations in VLC channels (i.e., fading) is mitigated since the area of a photo-diode is much larger than the light wavelength~\cite[Sec. 2.5]{dimitrov2015principles}. Thus, VLC channels are known as time-invariant. This fact is almost true regardless of the frame duration or the user being stationary or mobile in indoor environments, because the users are either stationary or move very slowly.}}}. Accordingly, the input-output relation in the VLC channel between the VLC access point of the transmitter and the photo-diode of the receiver at time instant $t$ is given as follows:
\begin{equation}\label{input_output_VLC}
	y_{v}(t)=\Omega x_{v}(t)g + w_{v}(t), 
\end{equation}
where $x_{v}(t)$ and $y_{v}(t)$ are the real-valued channel input and output, respectively. Above, $\Omega$ is the optical-to-electrical conversion efficiency (or detector responsivity) of the photo-diode in amperes per watt and $w_{v}(t)$ is the additive noise at the photo-diode of the receiver, which is a zero-mean, real Gaussian random variable with variance $\sigma_{v}^2$, i.e., $\mathbb{E}\{w_{v}^2\}=\sigma_{v}^2$. The noise samples $\{w_{v}\}$ are independent and identically distributed. Moreover, $g$ is the time-invariant optical channel gain.

Recall that the data transmitted over the VLC link is modulated on the light that illuminates the environment. Hence, assuming that the operation range of the radiated optical power is limited between $P_{\min}$ and $P_{\max}$ when the light is on, we modulate the data between the power levels $P_{\min}$ and $P_{\max}$, i.e., $P_{\min}\leq x_{v}(t) \leq P_{\max}$. As a result, the data bearing symbol, $\tilde{x}_{v}=x_{v}(t)-P_{\min}$ is limited as follows: $\tilde{x}_{v}\leq P_{\max}-P_{\min}= P_{\text{peak},v}$. Hence, we can re-express the input-output relation in \eqref{input_output_VLC} as
\begin{equation}\label{input_output_VLC_new}
	\tilde{y}_{v}(t)=y_{v}(t)-\Omega P_{\min}g=\Omega\tilde{x}_{v}(t)g + w_{v}(t).
\end{equation}
Now, assuming that the expected value of $\tilde{x}_{v}(t)$ is bounded by $P_{\text{avg},v}$, i.e., $\mathbb{E}\{\tilde{x}_{v}(t)\}\leq P_{\text{avg},v}$, and that the available bandwidth in the optical channel is $B_{v}$ Hz, we set the transmission rate in the channel in bits per frame to the lower bound on the channel capacity, which is defined as follows~\cite{lapidoth2009capacity}:
\begin{align}
	V&=\frac{TB_{v}}{2}\log_{2}\left\{1+P_{\text{peak},v}^2\frac{\Omega^2g^2}{2\pi\sigma_{v}^2}\exp\left\{2\frac{P_{\text{avg},v}}{P_{\text{peak},v}}\mu^{\star}-1\right\}\left(\frac{1-e^{-\mu^{\star}}}{\mu^{\star}}\right)^2\right\}\leq C_{v}, \label{V_l_1}
	\intertext{when $0\leq\frac{P_{\text{avg},v}}{P_{\text{peak},v}}<\frac{1}{2}$, and}
	V&=\frac{TB_{v}}{2}\log_{2}\left\{1+P_{\text{peak},v}^2\frac{\omega^2g^2}{2\pi\sigma_{v}^2}e^{-1}\right\}\leq C_{v}\label{V_l_2},
\end{align}
when $\frac{1}{2}\leq\frac{P_{\text{avg},v}}{P_{\text{peak},v}}\leq1$, where $\mu^{\star}$ is the unique solution to 
\begin{equation}\label{eq:mu}
	\frac{P_{\text{avg},v}}{P_{\text{peak},v}}=\frac{1}{\mu}-\frac{e^{-\mu}}{1-e^{-\mu}}.
\end{equation}
Above, $C_{v}$ and $V$ are constant values, because the mutual information in the VLC link does not change by time, i.e., due to the strong line-of-sight channel component, the channel gain, $g$, does not change. As for the input distribution, we refer to~\cite[Eq. 42]{lapidoth2009capacity}. 
%\vspace{-0.2cm}
\subsection{Source Model}\label{sec:source_model}
\begin{figure}
	\centering
	\includegraphics[width=0.3\columnwidth]{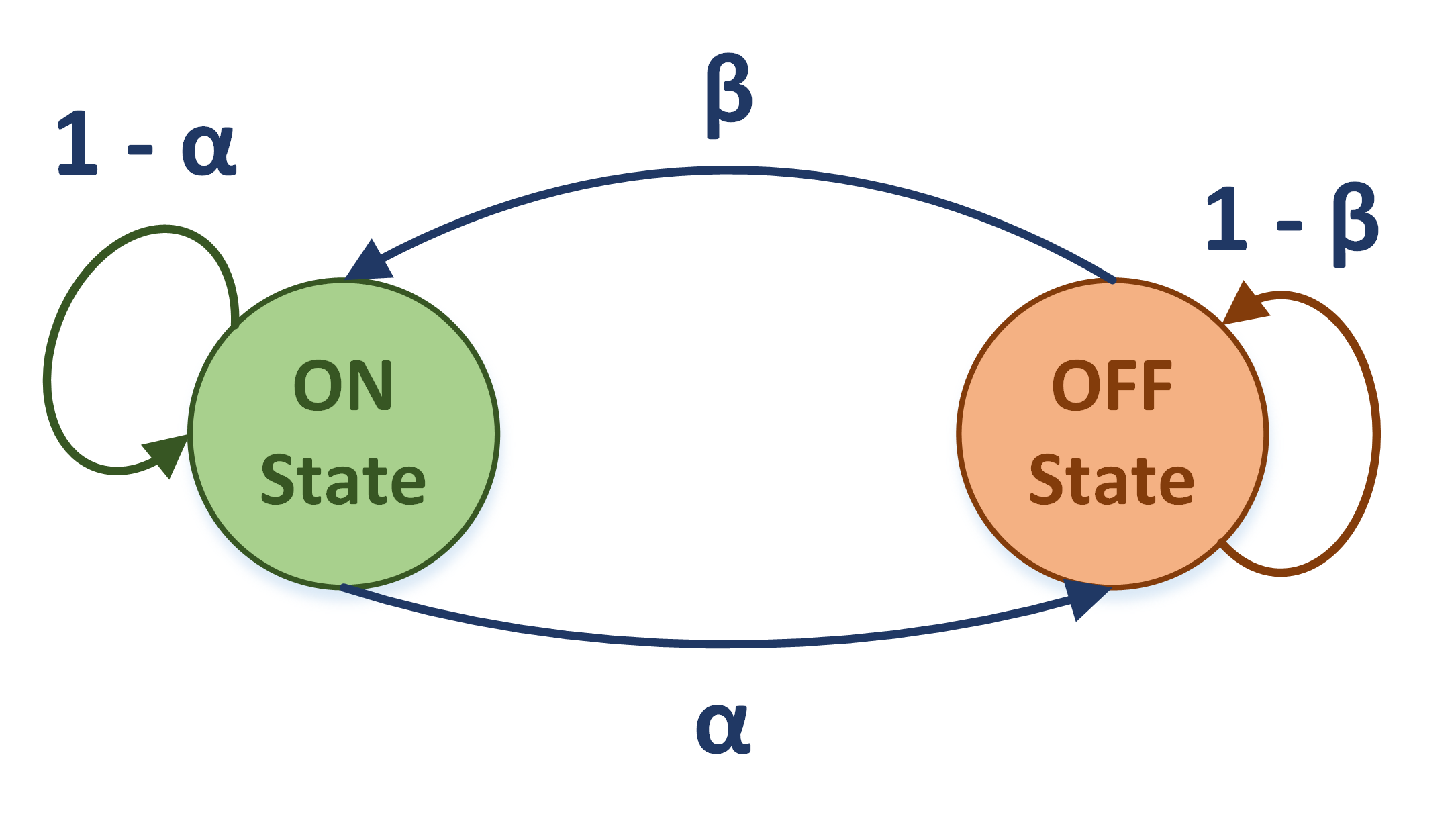}
	\caption{State transition model of the data arrival process.}\label{Source_Model}
	\vspace{-0.8cm}
\end{figure}
Regarding the data arrival process at the transmitter buffer, we consider a two-state discrete-time Markov process\footnote{We can project certain data arrival models on ON-OFF Markov processes. For instance, voice sources are generally modeled with ON and OFF states \cite{heffes1986markov}.} with ON and OFF states in each time frame. When the source is in the ON state in one time frame, the data from a source (or sources) arrives at the transmitter buffer. In the ON state, we consider a constant data arrival rate, i.e., $\lambda$ bits per frame. The number of bits arriving at the transmitter buffer is zero in the OFF state. As shown in Fig. \ref{Source_Model}, the transition probability from the ON state to the OFF state is denoted by $\alpha$ and the transition from the OFF state to the ON state is denoted by $\beta$. The probability of staying in the ON state is $1-\alpha$ and the probability of staying in the OFF state is $1-\beta$. Hence, the state transition matrix becomes
\begin{equation}
	J=\left[\begin{array}{cc}
		1-\alpha & \alpha \\
		\beta & 1-\beta
		\end{array}\right].
\end{equation}
Now, let $p_{\text{ON}}$ and $p_{\text{OFF}}$ be the steady-state probabilities of the data arrival process being in the ON and OFF states, respectively, where $p_{\text{ON}}+p_{\text{OFF}}=1$. Then, we have the following equality: $[p_{\text{ON}}\text{ }p_{\text{OFF}}]=[p_{\text{ON}}\text{ }p_{\text{OFF}}]J$. Subsequently, we have $p_{\text{ON}}=\frac{\beta}{\alpha+\beta}$ and $p_{\text{OFF}}=\frac{\alpha}{\alpha+\beta}$, and hence, the average data arrival rate at the transmitter buffer is $\frac{\beta\lambda}{\alpha+\beta}$ bits per frame. We finally note that the following analysis can easily be extended to other source models.

\section{Performance Analysis}\label{sec:Performance_Analysis}
In this section, we investigate the performance levels that the aforementioned system achieves by opportunistically exploiting the RF and VLC channels for data transmission. Herein, because the data is initially stored in the transmitter buffer before transmission, we assume that certain constraints are applied on the amount of the data in the buffer and the buffering delays. Therefore, we examine the system under QoS constraints that are associated with buffer overflow (data backlog) and buffering delay, and express the decay rate of the tail distribution of the queue length as~\cite[Eq. (63)]{chang1994stability}
\begin{equation}\label{buffer_limit}
	\theta=-\lim_{q \to \infty} \frac{\log_{\text{e}}\Pr\{Q\geq q\}}{q},
\end{equation}
where $Q$ is the stationary queue length, $Q(t)$ (see Fig. \ref{Channel_Model}), and $q$ is the buffer overflow threshold. Above, $\theta>0$ denotes the decay rate of the tail distribution of the data backlog, $Q$. Accordingly, we can approximate the buffer overflow probability for a large threshold, $q_{\max}$, as $\Pr\{Q\geq q_{\max}\}\approx e^{-\theta q_{\max}}$. Notice that the buffer overflow probability decays exponentially with a rate controlled by $\theta$, which is also defined as the QoS exponent. Basically, larger $\theta$ implies stricter QoS constraints, whereas smaller $\theta$ corresponds to looser constraints.

Recall that the outgoing service from the transmitter queue is $R_{l}$, given in \eqref{rate_set_lower_bound}, when the data is sent through the RF channel and it is $V$, given in \eqref{V_l_1} and \eqref{V_l_2}, when the data is sent through the VLC channel, while the data arrival is ON-OFF Markov process with $\lambda$ bits per frame in the ON state and zero bits in the OFF state. Hence, assuming that the buffer size is infinite and considering independent data arrival and work-conserving data service processes, there exist unique $\theta^{\star}_{r}>0$ and $\theta^{\star}_{v}>0$ such that $\Lambda_{a}(\theta^{\star}_{r})=-\Lambda_{r}(-\theta^{\star}_{r})$ and $\Lambda_{a}(\theta^{\star}_{v})=-\Lambda_{v}(-\theta^{\star}_{v})$, respectively, where $\Lambda_{a}(\theta)$, $\Lambda_{r}(\theta)$ and $\Lambda_{v}(\theta)$ are the asymptotic log-moment generating functions of the total amount of bits arriving at the transmitter buffer, the total service from the transmitter in the RF channel and the total service from the transmitter in the VLC channel, respectively~\cite[Theorem 2.1]{chang1995effective}. In particular, the asymptotic log-moment generating functions for any $\theta$ are
\begin{align}
\Lambda_{a}(\theta)&=\log_{\text{e}}\left\{\frac{1-\beta+(1-\alpha)e^{\theta\lambda}+\sqrt{\left[1-\beta+(1-\alpha)e^{\theta\lambda}\right]^2-4(1-\alpha-\beta)e^{\theta\lambda}}}{2}\right\},\label{eq:Lambda_A}\\
\Lambda_{r}(\theta)&=\log_{\text{e}}\left\{\mathbb{E}_{h}\left\{e^{\theta R_{l}}\right\}\right\}\text{ and }\Lambda_{v}(\theta)=\theta V.
\end{align}
We refer to \cite[Example 7.2.7]{chang2000performance} for obtaining the log-moment generating functions. Using $\Lambda_{a}(\theta^{\star}_{r})=-\Lambda_{r}(-\theta^{\star}_{r})$, we can express the maximum average data arrival rate at the transmitter buffer that the service process in the RF channel can sustain for any $\theta>0$ as
\begin{align}\label{eq:rho_r}
\rho_{r}(\theta)&=\frac{\beta}{(\alpha+\beta)\theta}\log_{\text{e}}\left\{\frac{e^{-2\Lambda_{r}(-\theta)}-(1-\beta)e^{-\Lambda_{r}(-\theta)}}{(1-\alpha)e^{-\Lambda_{r}(-\theta)}-(1-\alpha-\beta)}\right\}\nonumber\\
&=\frac{\beta}{(\alpha+\beta)\theta}\log_{\text{e}}\left\{\frac{1-(1-\beta)\mathbb{E}_{h}\left\{e^{-\theta R_{l}}\right\}}{(1-\alpha)\mathbb{E}_{h}\left\{e^{-\theta R_{l}}\right\}-(1-\alpha-\beta)\mathbb{E}^{2}_{h}\left\{e^{-\theta R_{l}}\right\}}\right\},
\end{align}
where $R_{l}$ is given in (\ref{rate_set_lower_bound}). \mar{For the derivation of $\rho_{r}(\theta)$, we refer to Appendix~\ref{app:der_rho_r}.} Likewise, using $\Lambda_{a}(\theta^{\star}_{v})=-\Lambda_{v}(-\theta^{\star}_{v})$ \mar{and following the steps in Appendix~\ref{app:der_rho_r}}, we can also express the maximum average data arrival rate at the transmitter buffer that the service process in the VLC channel can sustain for any $\theta>0$ as
\begin{align}\label{eq:rho_v}
	\rho_{v}(\theta)&=\frac{\beta}{(\alpha+\beta)\theta}\log_{\text{e}}\left\{\frac{e^{-2\Lambda_{v}(-\theta)}-(1-\beta)e^{-\Lambda_{v}(-\theta)}}{(1-\alpha)e^{-\Lambda_{v}(-\theta)}-(1-\alpha-\beta)}\right\}\nonumber\\ &=\frac{\beta}{(\alpha+\beta)\theta}\log_{\text{e}}\left\{\frac{e^{2\theta V}-(1-\beta)e^{\theta V}}{(1-\alpha)e^{\theta V}-(1-\alpha-\beta)}\right\},
\end{align}
where $V$ is given in \eqref{V_l_1} and \eqref{V_l_2} accordingly with the relation between $P_{\text{avg}}$ and $P_{\text{peak}}$. Moreover, in the special case where $\alpha=0$ and $\beta=1$, the expressions in \eqref{eq:rho_r} and \eqref{eq:rho_v} provide the effective capacity, which is the maximum sustainable constant data arrival rate by the channel process given the QoS constraints \cite{wu2003effective}. In another special case \mar{where $\alpha+\beta=1$, i.e., the state-transitions are independent of the past and current states}, we have
\begin{align}
\rho_{r}(\theta)=\frac{\beta}{\theta}\log_{\text{e}}\left\{\frac{1-\alpha\mathbb{E}_{h}\left\{e^{-\theta R_{l}}\right\}}{\beta\mathbb{E}_{h}\left\{e^{-\theta R_{l}}\right\}}\right\}\quad\text{and}\quad\rho_{v}(\theta)=\frac{\beta}{\theta}\log_{\text{e}}\left\{\frac{e^{2\theta V_{l}}-\alpha e^{\theta V_{l}}}{\beta e^{\theta V_{l}}}\right\}.
\end{align}

\subsection{Link Selection Policy}\label{sec:optimal}
In this section, we focus on the channel selection process that the transmitter employs. We set the maximum average data arrival rate under QoS constraints as the objective in the channel selection process. In particular, the transmitter chooses the channel in which the service process maximizes the average data arrival rate at the transmitter buffer. Notice that the transmission rate in the VLC channel is constant, whereas the transmission rate in the RF channel varies due to the changes in the channel fading gain. Now, due to the fact that the channel fading gains are known by the receiver as well as the transmitter, we provide the following proposition.

\begin{prop}\label{prob:prob_v}
In the aforementioned RF/VLC system, the transmitter sends data to the receiver over the VLC link when the following condition for a given QoS exponent, $\theta$, holds:
\begin{equation}\label{eq:comp}
V\geq\frac{1}{\theta}\log_{\text{e}}\left\{\frac{1-\beta+(1-\alpha)\xi+\sqrt{\left[1-\beta+(1-\alpha)\xi\right]^2-4(1-\alpha-\beta)\xi}}{2}\right\},
\end{equation}
where
\begin{equation}
\xi=\frac{1-(1-\beta)\mathbb{E}_{h}\left\{e^{-\theta R_{l}}\right\}}{(1-\alpha)\mathbb{E}_{h}\left\{e^{-\theta R_{l}}\right\}-(1-\alpha-\beta)\mathbb{E}^{2}_{h}\left\{e^{-\theta R_{l}}\right\}}.
\end{equation}
\end{prop}

\emph{Proof:} See Appendix \ref{app:prop_1}.$\hfill\square$

Proposition \ref{prob:prob_v} states that if the maximum attainable transmission rate in the VLC channel is greater than the right-hand side of \eqref{eq:comp}, the transmitter should perform transmission over the VLC link because the statistical variations in the RF channel deteriorates the buffer stability. Meanwhile, in the special case when $\alpha+\beta=1$, we re-express \eqref{eq:comp} as
\begin{equation}\label{eq:comp_add_1}
	V\geq-\frac{1}{\theta}\log_{\text{e}}\left\{\mathbb{E}_{h}\left\{e^{-\theta R_{l}}\right\}\right\}.
\end{equation}
Specifically, the constant rate in the VLC channel should be greater than the effective capacity of the RF channel such that the VLC channel is chosen for data transmission. In the following, we present two transmission strategies, such that (\textit{i}) the data is transmitted over the link with the highest instantaneous transmission rate in one time frame, and (\textit{ii}) the data is transmitted over both links simultaneously.

\paragraph*{Hybrid-Type I Transmission Strategy}\label{rem:handover}
In the above analysis, we obtain the performance levels when the transmitter chooses either of these two channels for data transmission following a link selection process based on the maximum average data arrival rates that the service processes in the channel can support. On the other hand, if there exists a fast and stable handover mechanism between the transmitter and the receiver, the transmitter will forward the data to the receiver over the link that provides the maximum lower bound on the instantaneous mutual information in the corresponding channel. For instance, when the lower bound on the instantaneous mutual information in the RF channel in the $l^{\text{th}}$ time frame, $R_{l}$, is greater than the lower bound on the instantaneous mutual information in the VLC channel, $V$, i.e., $R_{l}>V$, the transmitter sends the data over the RF link in the corresponding time frame only. Otherwise, it prefers sending the data over the VLC link. Respectively, we can establish the channel selection criterion as follows: The transmitter sends the data over the RF link when
\begin{equation}\label{eq:criterion_2}
	|h_{l}|^{2}>\frac{a\sigma_{r}^{2}}{2}\left(2^{\frac{V}{TB_{r}}}-1\right)\exp\left\{1-\frac{bP_{\text{avg},r}}{2}\right\}=\kappa.
\end{equation}
Otherwise, it sends the data over the VLC link. \mar{The aforementioned selection test can also be considered as the outage condition in the RF channel, i.e., the RF link is in outage when $|h_{l}|^{2}\leq\kappa$.} Noting that the channel fading gain changes independently from one time frame to another in the RF channel, the log-moment generating function of the service process becomes
\begin{equation}\label{eq:lambda_rv}
\Lambda_{rv}(\theta)=\log_{\text{e}}\left\{\mathbb{E}_{|h|^{2}>\kappa}\left\{e^{\theta R_{l}}\right\}+\Pr\{|h_{l}|^2\leq\kappa\}e^{\theta V}\right\},
\end{equation}
where $\mathbb{E}_{|h|^2>\kappa}\{e^{\theta R_{l}}\}$ is the conditional expectation given that $|h|^2>\kappa$, i.e., $\mathbb{E}_{|h|^2>\kappa}\{e^{\theta R_{l}}\}=\frac{1}{\Pr\{|h_{l}|^{2}>\kappa\}}\int_{\kappa}^{\infty}e^{\theta R_{l}}f_{|h|^2}(|h|^2)d|h|^2$, where $f_{|h|^2}(|h|^2)$ is the probability density function of $|h|^2$. Hence, the maximum average data arrival rate at the transmitter buffer that the hybrid service process can sustain under the QoS constraints specified by $\theta>0$ becomes
\begin{align}\label{eq:rho_rv}
\rho_{rv}(\theta)=\frac{\beta}{(\alpha+\beta)\theta}\log_{\text{e}}\left\{\frac{1-(1-\beta)D}{(1-\alpha)D-(1-\alpha-\beta)D^2}\right\},
\end{align}
where $D=\mathbb{E}_{|h|^{2}>\kappa}\left\{e^{-\theta R_{l}}\right\}+\Pr\{|h_{l}|^2\leq\kappa\}e^{-\theta V}$.

\paragraph*{Hybrid-Type II Transmission Strategy}\label{rem_2}
Different than the aforementioned protocols, we consider a transmitter that sends data over both links simultaneously in each frame. We assume that the receiver has a multihoming capability. We further assume a multiplexing-based transmission scheme such that the data streams transmitted over the RF and VLC links are different and independent from each other. Indeed, this scenario is feasible since the light and RF waves do not cause interference on each other. We further assume a power allocation policy between the two links, i.e., the average power constraint in the RF link is set to $P_{\text{avg},r}=\gamma_l P_{\text{avg}}$, and the one in the VLC link is set to $P_{\text{avg},v}=(1-\gamma_l)P_{\text{avg}}$, where $P_{\text{avg}}$ is the total average power constraint, and $0 < \gamma_l < 1$. Then, the instantaneous transmission rate in the RF channel in the $l^{\text{th}}$time frame becomes 
\begin{subequations}\label{rate_set_lower_bound_sim}
	\begin{equation}
		R_{l}=TB_r\log_{2}\left\{1+\frac{2|h_{l}|^2}{a\sigma_{r}^2}\exp\left\{\frac{b \gamma_l P_{\text{avg}}}{2}-1\right\}\right\}\quad\textrm{bits per frame}, \tag{\ref{rate_set_lower_bound_sim}}
	\end{equation}
	where $a$ and $b$ are the solutions of \eqref{eq:a_a} and \eqref{eq:a_b}.
\end{subequations}
Similarly following \eqref{V_l_1}--\eqref{V_l_2}, the transmission rate in the VLC channel becomes 
\begin{align}
	V&=\frac{TB_{v}}{2}\log_{2}\left\{1+P_{\text{peak},v}^2\frac{\Omega^2g^2}{2\pi\sigma_{v}^2}\exp\left\{2\frac{(1-\gamma_l)P_{\text{avg}}}{P_{\text{peak},v}}\mu^{\star}-1\right\}\left(\frac{1-e^{-\mu^{\star}}}{\mu^{\star}}\right)^2\right\}\label{V_l_1_sim},
	\intertext{when $0\leq\frac{(1-\gamma_l) P_{\text{avg}}}{P_{\text{peak},v}}<\frac{1}{2}$, and}
	V&=\frac{TB_{v}}{2}\log_{2}\left\{1+P_{\text{peak},v}^2\frac{\Omega^2g^2}{2\pi\sigma_{v}^2}e^{-1}\right\}\label{V_l_2_sim},
\end{align}
when $\frac{1}{2}\leq\frac{(1-\gamma_l) P_{\text{avg}}}{P_{\text{peak},v}}\leq1$, where $\mu^{\star}$ is the unique solution of \eqref{eq:mu}. It follows that the total transmission rate in each frame is the sum of the transmission rates in both links, i.e., $R_{l} + V$. Then, the log-moment generating function can be readily expressed as 
\begin{equation}\label{eq:24}
	\Lambda_{\text{srv}}(\theta) = \log_{\text{e}} \mathbb{E} \{e^{\theta (R_{l} + V)}\} = \theta V+\log_{\text{e}} \mathbb{E}\{e^{\theta R_{l}}\},
\end{equation}
and the maximum average data arrival rate is equal to
\begin{align}\label{eq:rho_sim}
	\rho_{\text{srv}}(\theta)=\frac{\beta}{(\alpha+\beta)\theta}\log_{\text{e}}\left\{\frac{1-(1-\beta)D}{(1-\alpha)D-(1-\alpha-\beta)D^2}\right\},
\end{align}
where $D= e^{-\theta V} \mathbb{E}_h \{e^{-\theta R_{l}}\}$. We finally remark that the optimal value of $\gamma_{l}$ that maximizes the sum transmission rate, i.e., $\text{max}_{\substack{\gamma_l}} \left\{R_{l} + V\right\}$, can be obtained numerically. 

\begin{rem}
Employing the aforementioned strategies requires the perfect knowledge of both RF and VLC channels at the transmitter side in each frame. \mar{We assume that the channel estimation is performed at the receiver, and forwarded to the transmitter in a delay-free and error-free feedback channel.} This increases the signaling overhead. Moreover, applying \emph{Hybrid-Type II Transmission Strategy} increases the implementation complexity because power sharing should be performed in each transmission frame. From the implementation perspective, exploiting \emph{Hybrid-Type II Transmission Strategy} is limited because the receiver should have a multihoming capability that enables it to perform link aggregation and receive data from different transmission technologies simultaneously. On the other hand, the maximum average data arrival rate is a steady-state performance measure. Thus, the selection process explained in Proposition~\ref{prob:prob_v} is considered as a large-timescale operation, which can be performed over periods of multiple transmission frames, and therefore, the implementation complexity decreases. Such a large-timescale operation is also proposed in~\cite{yu2016power}.
\end{rem}

\begin{rem}\label{rem:multi_VLC}
\mar{The selection process can also be employed in cases the access point controller has to choose between more than two transmission links. For example, let there be $N$ transmission links, and let \{$\rho_{1}(\theta),\dots, \rho_{N}(\theta)$\} be the maximum average data arrival rates at the transmitter buffer sustained by these links. Then, Proposition 1 in the paper will be updated with the solution of the following maximization problem:}
\begin{equation}
	\mar{\text{Transmission Link}=\max\left\{i:\rho_{i}(\theta)\right\}.}
\end{equation}
\mar{Similarly, let \{$R_{1}(l),\cdots,R_{N}(l)$\} be the instantaneous transmission rates provided by each link in the $l^{\text{th}}$ time frame. Then, the link selection criterion in \emph{Hybrid-Type I Transmission Strategy} will be updated with the solution of the following maximization problem:}
\begin{equation}
	\mar{\text{Transmission Link}=\max\left\{i:R_{i}\right\}.}
\end{equation}
\mar{Subsequently, the log-moment generating function of the service process will be}
\begin{equation}
	\mar{\Lambda_{1, \dots, N}(\theta)=\log_{\text{e}}\left\{{\Pr}_{i}\mathbb{E}{e^{\theta R_{i}(l)}}\right\},}
\end{equation}
\mar{where $\Pr_{i}$ is the probability that the transmitter chooses link $i$ for data transmission. Finally, if the receiver has a multihoming capability, \textit{Hybrid-Type II Transmission Strategy} can be applied with power sharing to maximize the total transmission rate in each frame.}
\end{rem}
%\vspace{-0.6cm}
\tcb{\subsection{Impacts of Handover Delay}\label{sec:handover}}
\tcb{Handover delay occurs when the transmitter moves from one link to the other, which is the case in \textit{Hybrid-Type I Transmission Strategy}. In this strategy, data transmission in one time frame is performed over the link that provides the maximum instantaneous transmission rate in that frame. Particularly, the transmitter switches from one link to the other or stay in the same link at the end of any time frame after comparing the instantaneous transmission rates in both links.}

\begin{figure}
	\centering
	\includegraphics[width=0.6\columnwidth]{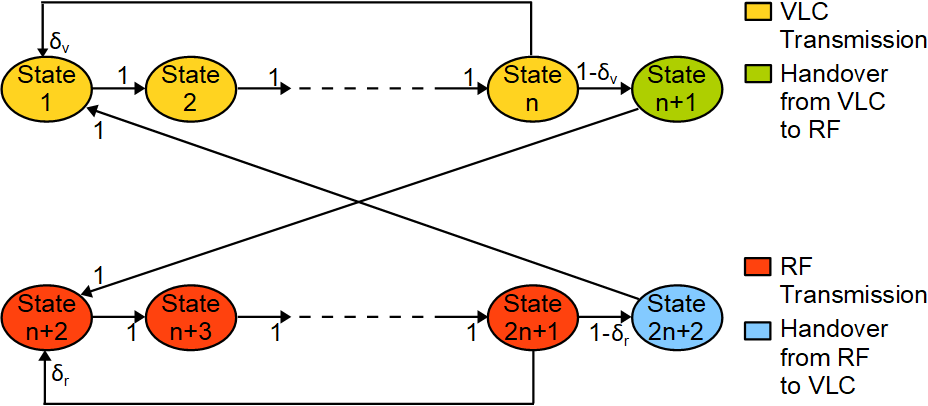}
	\caption{State transition model of the hybrid scenario with handover.}\label{fig:handover_approachI_noBlockage}
	\vspace{-0.8cm}
\end{figure}

\tcb{Now, given that $T_{\text{H}}$ denotes the duration of one single handover phase, let us initially assume that the frame duration is larger than the handover phase, i.e., $T> T_{\text{H}}$. For the sake of simplicity, we divide the frame duration into $n$ sub-frames that are equal to $T_{\text{H}}$, i.e., $T=n\times T_{\text{H}}$, where $n \in \mathbb{N}$ and $n > 1$. Particularly, a series of $n$ sub-frames of data transmission phase is followed by one sub-frame of handover process if the transmitter changes the transmission link, or by another series of $n$ sub-frames of data transmission phase if the transmitter stays in the same transmission link. For an analytical representation, we model the buffer activity at the end of each sub-frame as a discrete-time Markov process. As shown in Fig. \ref{fig:handover_approachI_noBlockage}, we have $2n+2$ states. The first $n$ states, i.e., \{State 1,$\cdots$,State $n$\}, represent the data transmission sub-frames in the VLC link, and State $(n+1)$ represents the handover process from the VLC link to the RF link. Similarly, the subsequent $n$ states, i.e., \{State $(n+2)$,$\cdots$,State $(2n+1)$\}, represent the data transmission sub-frames in the RF link, and State $(2n+2)$ represents the handover process from the RF link to the VLC link. Notice that the state transition probability from State $i$ to State $i+1$ is 1 for $i\in\{1,\cdots,n-1\}$ and for $i\in\{n+2,\cdots,2n\}$ because the data transmission in each link is completed in $n$ sub-frames and the link change may occur at the end of the $n^{\text{th}}$ and $(2n+1)^{\text{th}}$ sub-frames. On the other hand, in State $n$, either the transmitter changes the link and the system enters State $n+1$ with probability $1-\delta_{\nu}$, or the transmitter stays in the same link and the system enters State 1 with probability $\delta_{\nu}$. Similarly, in State $(2n+1)$, either the transmitter changes the link and the system enters State $(2n+2)$ with probability $1-\delta_{\text{r}}$, or the transmitter stays in the same link and the system enters State $(n+2)$ with probability $\delta_{\text{r}}$. Finally, the system moves from State $(n+1)$ to State $(n+2)$ and from State $(2n+2)$ to State 1 with probability 1 because at the end of one handover phase, the transmitter starts data transmission.}

\tcb{As seen in (\ref{eq:criterion_2}), if $|h_{l}|^{2}>\kappa$, the transmitter sends the data over the RF link. Otherwise, it sends the data over the VLC link. Therefore, for this specific case, we have $\delta_r = \text{Pr}\{|h_{l}|^{2}> \kappa\} = \delta$ and $\delta_v = 1 - \delta$. Then, we can express the $(2n+2)\times(2n+2)$ transition matrix as 
\begin{equation}\label{state_matrix}
	\Gamma = [p_{ji}], \quad \text{where} \quad p_{ji} = \begin{cases} 1, & \text{for} \; j = i+1 \quad \text{and} \; 1 \leq i \leq n-1 \; \text{or} \; n+2 \leq i \leq 2n,\\
	1, & \text{for} \; (i,j) = (n+1,n+2) \; \text{or} \; (i,j) = (2n+2,1),\\
	\delta, & \text{for} \; (i,j) = (n,n+1) \; \text{or} \; (i,j) = (2n+1,n+2),\\
	1-\delta, & \text{for} \; (i,j) = (n,1) \; \text{or} \; (i,j) = (2n+1,2n+2),\\
	0, & \text{otherwise,}\end{cases} 
\end{equation}
where $p_{ji}$ is the state transition probability from State $i$ to State $j$. Then, we can re-characterize the log-moment generating function of the hybrid system for any $\theta>0$, which is provided in (\ref{eq:lambda_rv}), as follows \cite[Chap. 7, Example 7.2.7]{chang2012performance}:
\begin{equation}\label{eq:mgf_hybrid}
	\Lambda_{rv}(\theta) = \log_{\text{e}}\spp(\Phi(\theta)\Gamma),
\end{equation}
where $\spp(\Phi(\theta)\Gamma)$ is the spectral radius of the matrix $\Phi(\theta)\Gamma$, and $\Phi(\theta)$ is a diagonal matrix whose components are the moment generating functions of the processes in $2n+2$ states. Notice that the transmitted bits are removed from the transmitter buffer only at the ends of the $n^{\text{th}}$ and $(2n+1)^{\text{th}}$ frames. Therefore, the moment generating functions are $e^{\theta V}$ and $\mathbb{E}_{|h|^2>\kappa}\{e^{\theta R_{l}}\}$ in the $n^{\text{th}}$ and $(2n+1)^{\text{th}}$ frames, respectively. However, there are no bits removed in the other states, i.e., the service rates in the other states are effectively zero. Hence, the moment generating functions are 1 in other states. Moreover, the unique QoS exponent, $\theta^{\star}$, is obtained when $\Lambda_{a}(\theta^{\star})=-n\Lambda_{rv}(-\theta^{\star})$.}

\subsection{Non-asymptotic Bounds}\label{sec:Non-asymptotic_Bounds}
The aforementioned results provide the performance analysis in the steady-state. Particularly, the analysis is obtained when the number of time frames is very large. On the other hand, non-asymptotic bounds regarding the statistical queueing and delay characterizations at the transmitter buffer are of interest for system designers as well. Therefore, we address the framework of Network Calculus~\cite{chang2000performance,fidler2015guide,ciucu2006scaling}, and consider~\cite[Theorem 2]{fidler2015capacity}, which states that a minimal bound on the queue length can be found for a given buffer overflow probability. Particularly, given the RF-based data service process in Section~\ref{sec:RF_Channel_Model} and the two-state Markov modeled data arrival process in Section \ref{sec:source_model}, the buffer threshold, $q$, is expressed as
\begin{equation}\label{eq:queue_overflow_bound}
	q=\inf_{c>0}\{q_{r}+q_{a}\},
\end{equation}
for a given buffer overflow probability, $\Pr\{Q\geq q\}=\varepsilon$, where
\begin{align}\label{eq:q_r}
	q_{r}=-\sup_{\theta}\left\{\frac{\log_{\text{e}}\left\{-\varepsilon_{r}\left[\Lambda_{r}(-\theta)+\theta c\right]\right\}}{\theta}\right\}\text{ for }\max\left\{0,-\frac{1}{c\varepsilon_{r}}-\frac{\Lambda_{r}(-\theta)}{c}\right\}< \theta,
\end{align}
and
\begin{align}\label{eq:q_a}
	q_{a}=-\sup_{\theta}\left\{\frac{\log_{\text{e}}\left\{\varepsilon_{a}\left[\theta c-\sup_{t>0}\left\{\Lambda_{a}\left(\theta,t\right)\right\}\right]\right\}}{\theta}\right\}\text{ for }0<\theta<\frac{1}{c\varepsilon_{a}}+\frac{\sup_{t>0}\left\{\Lambda_{a}\left(\theta,t\right)\right\}}{c},
\end{align}
with
\begin{align}
	\Lambda_{a}\left(\theta,t\right)=\frac{1}{t}\log_{\text{e}}\left\{[\begin{matrix}p_{\text{ON}}&p_{\text{OFF}}\end{matrix}]\left(\left[\begin{matrix}e^{\theta\lambda}&0\\0&1\end{matrix}\right]\left[\begin{matrix}1-\alpha&\alpha\\\beta&1-\beta\end{matrix}\right]\right)^{(t-1)}\left[\begin{matrix}e^{\theta\lambda}&0\\0&1\end{matrix}\right]\left[\begin{matrix}1\\1\end{matrix}\right]\right\}.
\end{align}
Above, the buffer violation probability is $\varepsilon=\varepsilon_{r}+\varepsilon_{a}$, and $\theta$ and $c$ are free parameters. Notice also that the log-moment generating function in (\ref{eq:q_r}), i.e., $\Lambda_{r}(-\theta)$, is time-invariant because the service process is memory-less, whereas the log-moment generating function in (\ref{eq:q_a}), i.e., $\Lambda_{a}\left(\theta,t\right)$, is time-variant because the arrival process depends on its current state. Moreover, we remark that when $t$ goes to infinity, we have $\lim_{t\to\infty}\Lambda_{a}\left(\theta,t\right)=\Lambda_{a}\left(\theta\right)$, where $\Lambda_{a}\left(\theta\right)$ is expressed in \eqref{eq:Lambda_A}. Herein, we refer to \cite{fidler2015capacity} for further calculation details in \eqref{eq:q_r} and \eqref{eq:q_a}. 

Likewise, when the VLC-based data service process in Section \ref{sec:VLC_Channel_Model} is employed, we have
\begin{equation}\label{eq:queue_overflow_bound_v}
	q=\inf_{c>0}\{q_{v}+q_{a}\},
\end{equation}
and
\begin{align}\label{eq:q_v}
	q_{v}=-\sup_{\theta}\frac{\log_{\text{e}}\left\{-\varepsilon_{r}\left[\Lambda_{v}(-\theta)+\theta c\right]\right\}}{\theta}=-\sup_{\theta}\frac{\log_{\text{e}}\left\{-\varepsilon_{r}\left[-\theta V+\theta c\right]\right\}}{\theta}.
\end{align}
Notice that $q$ in \eqref{eq:queue_overflow_bound_v} is minimized when $c=V$. Moreover, because $q_{v}$ cannot be smaller than zero, we have $q=q_{a}$ and $c=V$. Therefore, when the VLC-based service channel is chosen as the service process, we have
\begin{align}
q=q_{a}=-\sup_{\theta}\left\{\frac{\log_{\text{e}}\left\{\varepsilon_{a}\left[\theta V-\sup_{t>0}\left\{\Lambda_{a}\left(\theta,t\right)\right\}\right]\right\}}{\theta}\right\}\text{ and }\varepsilon=\varepsilon_{a}.
\end{align}

Now, assuming a fast and stable handover mechanism between the transmitter and the receiver, and a service channel selection process as described in \emph{Hybrid-Type I Transmission Strategy}, we can characterize the delay bound as follows: $q=\inf_{c>0}\{q_{a}+q_{rv}\}$ where
\begin{align}
	q_{rv}=-\sup_{\theta}\left\{\frac{\log_{\text{e}}\left\{-\varepsilon_{r}\left[\Lambda_{rv}(-\theta)+\theta c\right]\right\}}{\theta}\right\}\text{ for }\max\left\{0,-\frac{1}{c\varepsilon_{r}}-\frac{\Lambda_{rv}(-\theta)}{c}\right\}< \theta,
\end{align}
and $\Lambda_{rv}(\theta)$ is given in (\ref{eq:lambda_rv}).

\begin{rem}\label{rem_3}
Let us assume a first-come first-served protocol exists at the transmitter buffer. Then, the minimal bound on the buffering delay is expressed as follows \cite[Theorem 1]{fidler2015capacity}:
\begin{equation}\label{eq:queue_delay_bound}
d=\inf_{c>0}\left\{\frac{q_{r}+q_{a}}{c}\right\},\text{ or }d=\inf_{c>0}\left\{\frac{q_{v}+q_{a}}{c}\right\},\text{ or }d=\inf_{c>0}\left\{\frac{q_{rv}+q_{a}}{c}\right\},
\end{equation}
when the RF-based service process, or the VLC-based service process, or \emph{Hybrid-Type I Transmission Strategy} is employed.
\end{rem}

\begin{rem}\label{rem:multi}
\mar{We consider a user-related performance measure, i.e., the maximum average data arrival rate at the data buffer, and formulate the link selection by employing the transmission rates provided by the RF and VLC links. Our analytical framework can easily be extended to a more general multi-user scenario by regarding the rate allocations for each user in the transmission links and the \emph{receiver-oriented} data arrival processes at the transmitter buffer. In this regard, we refer to Fig. \ref{Rate_RF_VLC_num_users} in Section \ref{sec:sim}, where we employ frequency-division multiple access (FDMA) and time-division multiple access (TDMA) protocols for numerical illustrations. Moreover, our paper is different than \cite{jin2015resource,wang2015dynamic,yu2016power,wu2017dynamic}. The system sum throughput is maximized in \cite{jin2015resource,wang2015dynamic}, and the system average power consumption is minimized in \cite{yu2016power,wu2017dynamic}. In these studies, a framework in which a joint resource allocation and link assignment process is employed is not provided, and the optimization problems are in principle mixed integer and non-linear programming problems, which are mathematically intractable. Therefore, the main optimization problems are decomposed into solvable sub-problems, and iterative algorithms are provided.}

\end{rem}
\begin{figure}
	\centering
	\includegraphics[width=0.3\columnwidth]{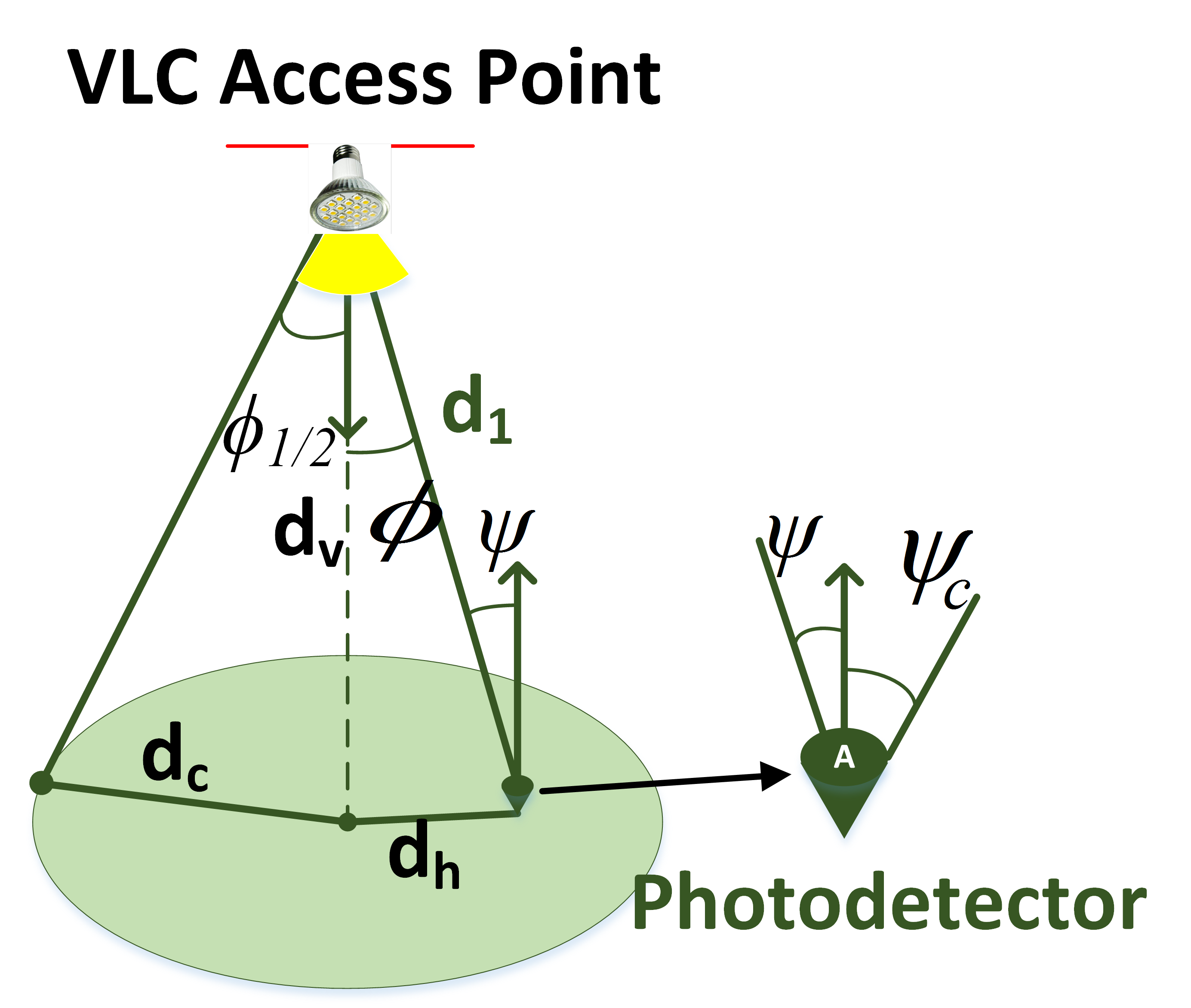}
	\caption{The VLC channel model.}\label{VLC_LOS}
	\vspace{-0.8cm}
\end{figure}

\begin{rem}\label{rem_user_mobility}
\tcb{The link selection process can easily be adopted into scenarios where the receiver is mobile. All that one needs is to consider the log-moment generating function in the VLC link for changing transmission rates, i.e., $\Lambda_{v}(\theta)=\log_{\text{e}}\left\{\mathbb{E}_{v}\{e^{\theta V_{l}}\}\right\}$, and base the channel selection process on the link that increases the maximum average data arrival rate at the transmitter buffer. In \emph{Hybrid-Type I Transmission Strategy} and \emph{Hybrid-Type II Transmission Strategy}, because the channel fading gains in the RF and VLC links are instantaneously known at the transmitter in each time frame, the aforementioned analysis will not be different than what we currently have in the paper even if the transmission rates vary due to mobility. We also note that the user mobility is normally low in indoor scenarios. In this regard, we refer to \cite{dehghani2015limited}.}
\end{rem}

\section{Numerical Results}\label{sec:sim}
\begin{table}[t]
\caption{Simulation Parameters}
\begin{center}
\begin{tabular}{c|c|c|c}\hline\hline
 \multicolumn{4}{c}{{\bf VLC System}}\\
\hline
LED half intensity viewing angle, $\phi_{1/2}$ & $\{30^{\circ},45^{\circ},60^{\circ}\}$ & 
PD field of view (FOV), $\psi_C$ & $ 90^{\circ}$ \\
PD physical area, $A$ & $ 1 \,\text{cm}^2$ &
Channel bandwidth, $B_{v}$ & $10$ MHz\\ 
PD opt.-to-elect. conversion efficiency, $\Omega$ & $0.53$ A/W &
PD optical concentrator gain, $u(\psi)$ & $1$\\ 
Vertical distance, $d_v$ & $2.5$ m &
Noise power spectral density, $N_{v}$ & $10^{-21}$ $\text{A}^2/$ Hz\\
\hline
 \multicolumn{4}{c}{{\bf RF System}}\\
\hline
Channel bandwidth, $B_{r}$ & $10$ MHz & 
Path loss exponent, $q$ & $1.8$ \\
Rician factor, $K$ & $10$ dB & 
Log-normal standard deviation, $\sigma$ & $3.6$ dB\\ 
Ambient Temperature, $T_t$ & $280^ \circ$ K & 
&\\
\hline\hline
\end{tabular}
\end{center}\label{tab_1}
\vspace{-0.6cm}
\end{table}
In this section, we present numerical results that substantiate our theoretical findings. Unless otherwise specified, we set the transmission time frame to $0.1$ milliseconds, i.e., $T=10^{-4}$. We assume that the LED at the VLC access point of the transmitter has a Lambertian radiation pattern, and that the transmitter and receiver planes are parallel to each other. We further assume that the transmitter is directed downwards, while the receiver is directed upwards, as depicted in \figurename~\ref{VLC_LOS}. However, our theoretical results can easily be adopted into different positional settings. Herein, the line-of-sight channel gain, i.e., $g$, is given as follows \cite{barry1993simulation}:
\begin{equation}\label{VLC_gain}
	g = \frac{(s+1)A u(\psi) d_v^{s+1}}{2 \pi d_{1}^{s+3}} \text{rect}(\psi/\psi_C),
\end{equation}
where $A$ is the surface area of the photo-diode, $d_{1}$ is the distance between the LED at the transmitter and the photo-diode at the receiver, and $d_v$ is the normal distance between the transmitting and receiving planes. Moreover, $u(\psi)$ is the optical concentrator gain at the photo-diode, $\psi$ is the angle of incidence, and $\psi_C$ is the field of view, i.e., the maximum angle at which the light emitted by the LED is detected by the photo-diode. In addition, $s=\frac{-1}{\log_2(\cos(\phi_{1/2}))}$ is the Lambertian index, where $\phi_{1/2}$ is the LED half intensity viewing angle, and $\text{rect}(x) = 1$ if $|x| \leq 1$ and $\text{rect}(x) = 0$ otherwise. Finally, the thermal noise power at the photo-diode is $\sigma_v^2=N_v B_v$, where $N_v$ is the noise power spectral density.

Regarding the RF channel, we consider a Rician fading distribution with the Rician factor, $K$, where the channel realizations, $\{h_l\}$, are independent and identically distributed, circularly symmetric complex Gaussian random variables with mean $\mu = \sqrt{\frac{e^{-L(d_0)/10} K}{K+1}}$ and variance $\sigma_h^2 = \frac{e^{-L(d_0)/10}}{K+1}$. \mar{Setting $K$ to a reasonable value, we can reflect the channel characteristics in millimeter wave range communications as well \cite{basnayaka2017design}.} Here, $L(d_0)$ is the large-scale path loss in {\it decibels} as a function of the distance between the RF access point at the transmitter and the RF front-end at the receiver, $d_0$, and it is given by \cite{akl2006indoor} 
\begin{equation}
L(d_0) = L ({d_{\text{ref}}}) + 10 q \log_{\text{e}}\bigg(\frac{d_{0}}{d_{\text{ref}}}\bigg) + X_{\sigma},
\end{equation}
 where $L ({d_{\text{ref}}}) = 40$ dB is the path loss at a reference distance, ${d_{\text{ref}}} = 1$ m, and an operating frequency of $2.4$ GHz. In addition, $q$ is the path loss exponent and $X_{\sigma}$ represents the shadowing effect, which is assumed to be a zero-mean Gaussian random variable with a standard deviation $\sigma$ expressed in decibels\footnote{Empirical values of $K$, $q$, and $\sigma^2$ in different indoor scenarios were provided in \cite{akl2006indoor,rappaport1989uhf, neskovic2000modern}. For instance, the value of $q$ ranges from $1.2$ to $1.7$, while $\sigma$ varies between $3.6$ dB and $4.0$ dB inside buildings \cite{akl2006indoor}.}. Finally, the thermal noise power at the RF front-end of the receiver is $\sigma_r^2 = \kappa_B T_t B_r$, where $\kappa_B$ is the Boltzmann constant and $T_t$ is the ambient temperature \cite{stefan2013area}. Table \ref{tab_1} summarizes the simulation parameters, unless otherwise stated. Finally, setting the average transmission power constraint to $P_{\text{avg}}$ in all the transmission strategies, we define $\nu$ as the average-to-peak power ratio in the RF and VLC links, i.e., $\nu = \frac{P_{\text{avg}}}{P_{\text{peak}, r}} = \frac{P_{\text{avg}}}{P_{\text{peak},v}}$.

\begin{figure}[t]
	\centering
	\subfigure[$\theta = 0.01$]{
	\includegraphics[width=\figsize\textwidth]{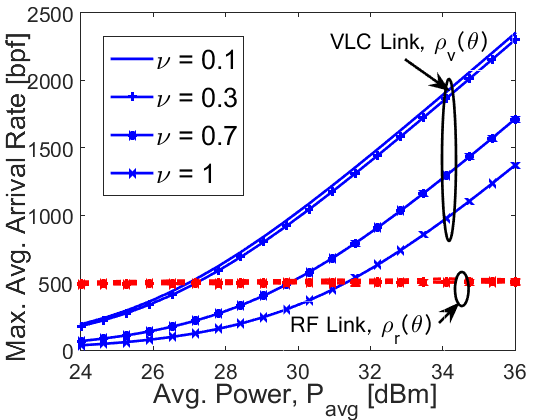}\label{Rate_RF_VLC_SNR_nu_theta_0_0_1}}
	\quad\quad
	\subfigure[$\theta = 0.1$]{
	\includegraphics[width=\figsize\textwidth]{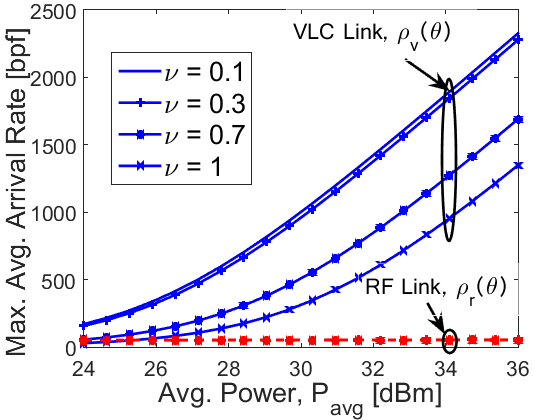}\label{Rate_RF_VLC_SNR_nu_theta_0_1}}
	\caption{Maximum average arrival rates of VLC and RF links as a function of the average power limit, $P_{\text{avg}}$, for different values of the average-to-peak power ratio $\nu$ and the QoS exponent, $\theta$. Here, $d_0 = 15$ m, $d_1 = 3$ m, $\phi_{1/2} = 60^{\circ}$, $\alpha = 0.3$, and $\beta = 0.7$. \{bpf : bits per frame\}.}\label{Rate_RF_VLC_SNR_nu_theta}
	\vspace{-0.8cm}
\end{figure}

\mar{\subsection{Transmission Strategies}}
We consider the scenario in which a transmitter has a VLC access point and an RF access point, as shown in \figurename~\ref{Channel_Model}. The receiver is located at a distance $d_1 = 3$ m from the VLC access point, \mar{i.e., the user is located at a horizontal distance $d_h \approx 1.6$ m from the cell center, where $\phi_{1/2} = 60^{\circ}$ and $d_v = 2.5$ m}, and at a distance $d_0 = 15$ m from the RF access point. In \figurename~\ref{Rate_RF_VLC_SNR_nu_theta}, we plot the maximum average data arrival rates at the transmitter buffer, $\rho_{v}(\theta)$ and $\rho_r(\theta)$, as functions of the average power constraint, $P_{\text{avg}}$, with different average-to-peak power ratios, i.e, $\nu \in \{0.1,0.3,0.7,1\}$, when the VLC and RF links are employed, respectively. We have the results for $\theta=0.01$ in \figurename~\ref{Rate_RF_VLC_SNR_nu_theta_0_0_1} and for $\theta = 0.1$ in \figurename~\ref{Rate_RF_VLC_SNR_nu_theta_0_1}. We observe that the maximum average data arrival rates increase faster with the increasing average power constraint in the VLC link than they do in the RF link. For instance, when $\nu = 0.3$ and $\theta = 0.01$, the maximum average data arrival rate increases more than $145$ bits per frame with $P_{\text{avg}}$ increasing from $27$ dBm to $28$ dBm in the VLC link, whereas it increases $2$ bits per frame in the RF link. We observe the same behavior when the peak power constraint increases. We can explain this result with the stochastic nature of the transmission rates in the RF channel. Particularly, when the instantaneous transmission rate in the RF channel becomes very low, more data packets are accumulated in the transmitter buffer. Therefore, in order to sustain the QoS constraints, the transmitter buffer should accept data at lower arrival rates. On the other hand, because the transmission rate in the VLC channel is constant, the maximum average data arrival rate increases almost linearly with the increasing transmission rate in the VLC channel. Moreover, the performance in the RF link is better than the VLC link when the average power constraint is lower, and the performance in the VLC link is better than the RF link when the average power constraint is higher.

\begin{figure}[t]
	\centering
	\subfigure[$P_{\text{avg}} = 24$ dBm]{
	\includegraphics[width=\figsize\textwidth]{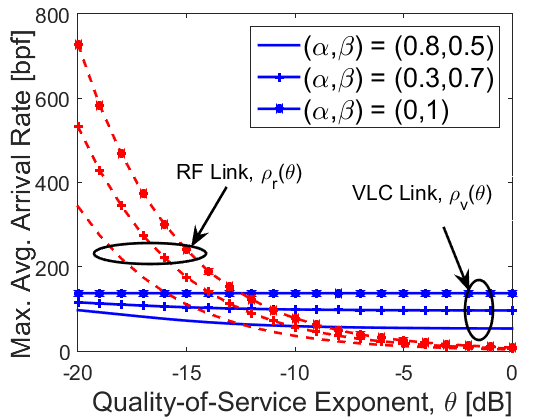}\label{Rate_RF_VLC_theta_alpha_beta_SNR_-10}}
	\quad\quad
	\subfigure[$P_{\text{avg}} = 30$ dBm]{
	\includegraphics[width=\figsize\textwidth]{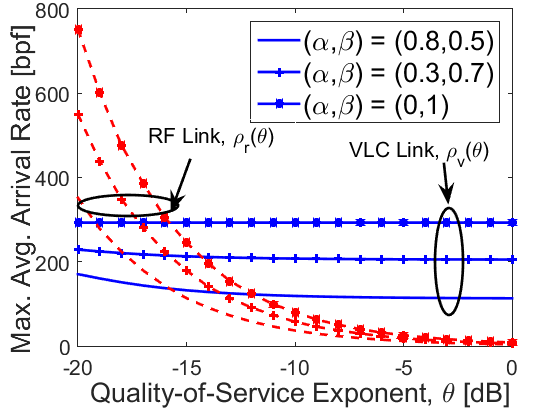}\label{Rate_RF_VLC_theta_alpha_beta_SNR_-5}}
	\caption{Maximum average arrival rates of VLC and RF links as a function of the QoS exponent, $\theta$, for different values of the average power limit $P_{\text{avg}}$ and the source statistics, $\alpha$ and $\beta$. Here, $d_0 = 15$ m, $d_1 = 3$ m, $\alpha = 0.3$, $\phi_{1/2} = 60^{\circ}$, and $\nu = 0.7$. \{bpf : bits per frame\}.}\label{Rate_RF_VLC_theta_alpha_beta_SNR}
	\vspace{-0.8cm}
\end{figure}

In \figurename~\ref{Rate_RF_VLC_theta_alpha_beta_SNR}, we plot $\rho_{r}(\theta)$ and $\rho_v(\theta)$ as functions of the QoS exponent, $\theta$, for different values of $\alpha$ and $\beta$ when $\nu = 0.7$. We have $P_{\text{avg}} = 24$ dBm in \figurename~\ref{Rate_RF_VLC_theta_alpha_beta_SNR_-10} and $P_{\text{avg}} = 30$ dBm in \figurename~\ref{Rate_RF_VLC_theta_alpha_beta_SNR_-5}. The performance in the RF link is better than the VLC link when $\theta$ is low, whereas the performance in the RF link decreases faster with increasing $\theta$ and becomes less than the performance in the VLC link. In other words, the stochastic nature of the RF channel prevents the RF link from supporting data arrival rates at the transmitter buffer when the QoS constraints are stringent. Indeed, the maximum average data arrival rate that the RF link supports approaches zero exponentially with increasing $\theta$ regardless of the average power constraint and the source statistics, i.e., $\alpha$ and $\beta$. However, the RF link can support higher data arrival rates if the QoS constraints are looser. We further observe that increasing $\beta$ (or decreasing $\alpha$), results in better performance values in the VLC and RF links because the steady-state probability of the ON state, i.e., $P_{\text{ON}} = \frac{\beta}{\alpha + \beta}$, and the average arrival rate at the transmitter buffer increase. However, the effect of the source statistics on the performance values is much less than that of the QoS constraints, especially in the RF link. In other words, the randomness in the service process has a higher impact on the system performance than the randomness in the arrival process has.

\begin{figure}[t]
	\centering
	\subfigure[FDMA (i.e., power and bandwidth)]{
	\includegraphics[width=\figsize\textwidth]{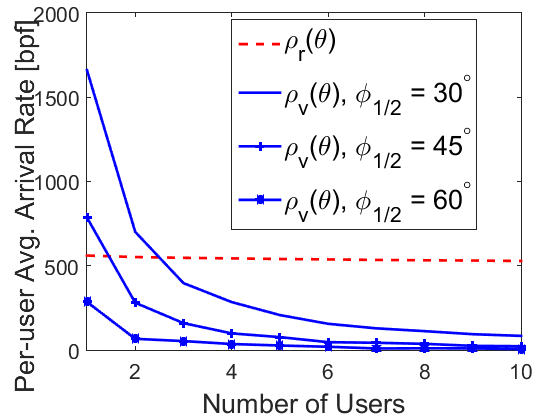}\label{Rate_RF_VLC_num_users_FDMA}}
	\quad\quad
	\subfigure[TDMA (i.e., time)]{
	\includegraphics[width=\figsize\textwidth]{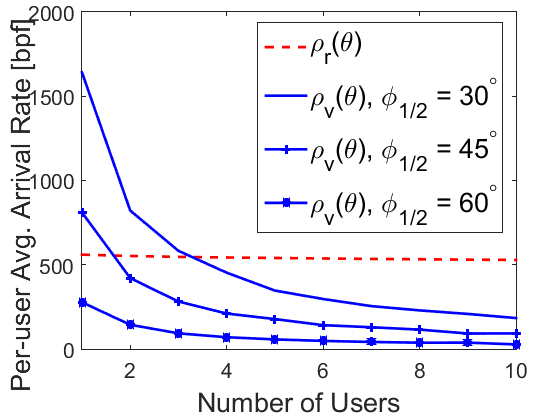}\label{Rate_RF_VLC_num_users_TDMA}}
	\caption{Per-user maximum average arrival rates of VLC and RF links as a function of the number of served receivers (or equivalently the receiver allocated resources) and for different values of the LED viewing angle, $\phi_{1/2}$. Here, $\alpha = 0.3$, $\beta = 0.7$, $\nu = 1$, and $P_{\text{avg}} = 24$ dBm. \{bpf : bits per frame\}.}\label{Rate_RF_VLC_num_users}
	\vspace{-0.8cm}
\end{figure}

In typical indoor scenarios, VLC and RF access points serve multiple receivers. This is applied by sharing available resources (i.e., power, time, and bandwidth) among the served receivers. Herein, we assume that the transmitter employs the commonly known FDMA and TDMA schemes in both links. \mar{In \figurename~\ref{Rate_RF_VLC_num_users}, we plot the maximum average data arrival rate per user given that all the users are uniformly positioned within the coverage area of the VLC access point.} We observe that the performance per user in the RF link is generally much higher than the performance per user in the VLC link when the number of receivers is above 4. Basically, the system can serve more users in the RF link than the VLC link when the QoS constraints are of interest. In addition, the results in \figurename~\ref{Rate_RF_VLC_num_users} agree with the results in \figurename~\ref{Rate_RF_VLC_SNR_nu_theta}, where the performance of the VLC link is highly affected by the decreasing average power constraint. We finally see that with the decreasing LED viewing angle, the performance in the VLC link becomes better because the transmission power is concentrated in smaller areas. Notice also that the VLC channel gain is affected by $\phi_{1/2}$ through the Lambertian index. \mar{Herein, we show the performance sensitivity of the RF and VLC links to the allocated transmission resources such as power, time, and bandwidth, given that the available resources are equally shared among the users. We also show that our framework can easily be invoked in a multi-user scenario.}

We further explore the system performance with respect to the receiver location. We set $(x_v, y_v, z_v) = (0, 0, 0)$ as the Cartesian coordinates of the VLC access point, $(x_r, y_r, z_r) = (10, 0, 0)$ as the coordinates of the RF access point, and $(x_u, y_u, z_u)$ as the coordinates of the receiver, where $z_u = -d_v$. Particularly, we consider the following strategies: 
\begin{enumerate}
	\item \emph{RF-only Strategy}: The transmitter sends data over the RF link only, and the maximum average arrival rate, $\rho_r(\theta)$, is expressed in \eqref{eq:rho_r}.
	\item \emph{VLC-only Strategy}: The transmitter sends data over the VLC link only, and the maximum average arrival rate, $\rho_v(\theta)$, is expressed in \eqref{eq:rho_v}. 
	\item \emph{Hybrid-type I Transmission Strategy}: The transmitter sends data over the link that provides the highest transmission rate, and the maximum average arrival rate, $\rho_{rv}(\theta)$, is expressed in \eqref{eq:rho_rv}.
	\item \emph{Hybrid-type II Transmission Strategy}: The transmitter sends data over both links simultaneously following a power allocation policy to maximize the transmission rate, and the maximum rate, $\rho_{\text{srv}}(\theta)$, is expressed in \eqref{eq:rho_sim}. 
\end{enumerate}
\begin{figure}
	\centering
	\subfigure[RF-only,, $\theta = 0.01$]{\includegraphics[width=\figsize\textwidth]{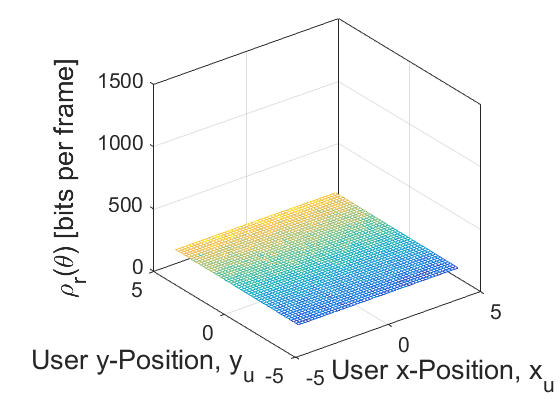}\label{max_avg_theta_60_path_2_a}}\quad\quad
	\subfigure[VLC-only, $\theta = 0.01$]{\includegraphics[width=\figsize\textwidth]{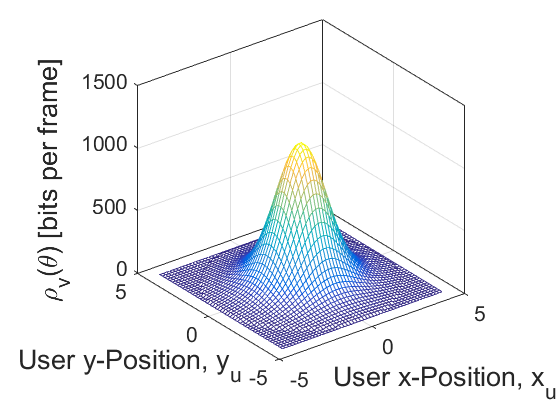}\label{max_avg_theta_60_path_1_a}}\\
	\subfigure[Hybrid-type I, $\theta = 0.01$]{\includegraphics[width=\figsize\textwidth]{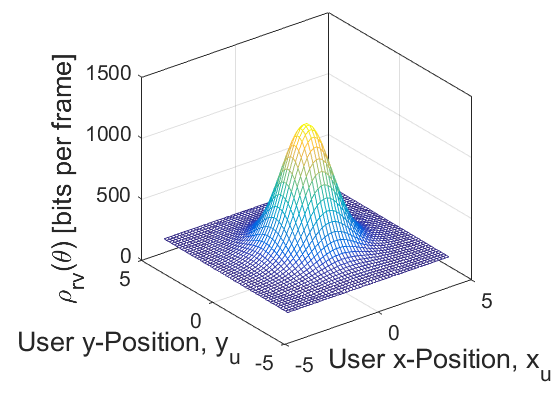}\label{max_avg_theta_60_path_1_theta_0_1_a}}\quad
	\subfigure[Hybrid-type II, $\theta = 0.01$]{\includegraphics[width=\figsize\textwidth]{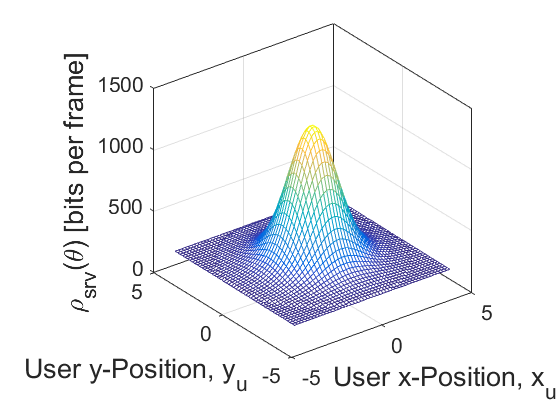}\label{max_avg_theta_60_path_2_theta_0_1_a}}\\
	\subfigure[RF-only,, $\theta = 0.1$]{\includegraphics[width=\figsize\textwidth]{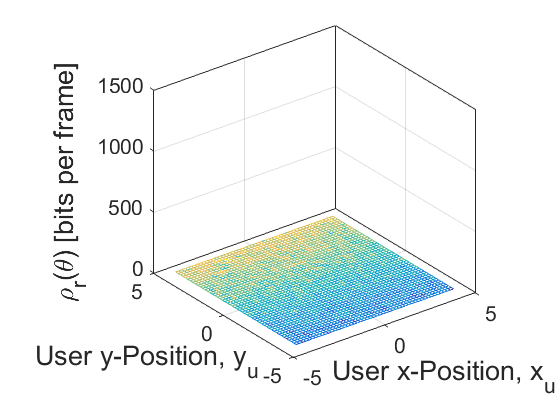}\label{max_avg_theta_60_path_2_b}}\quad
	\subfigure[VLC-only, $\theta = 0.1$]{\includegraphics[width=\figsize\textwidth]{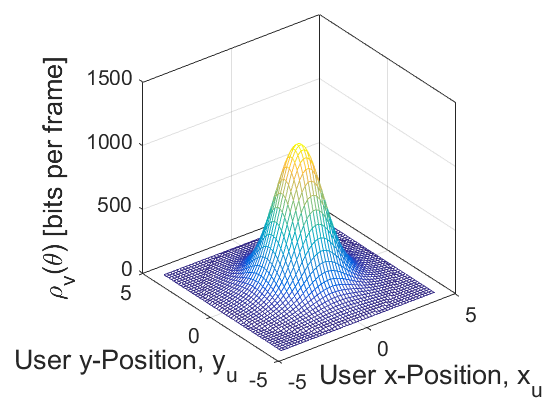}\label{max_avg_theta_60_path_1_b}}\\
	\subfigure[Hybrid-type I, $\theta = 0.1$]{\includegraphics[width=\figsize\textwidth]{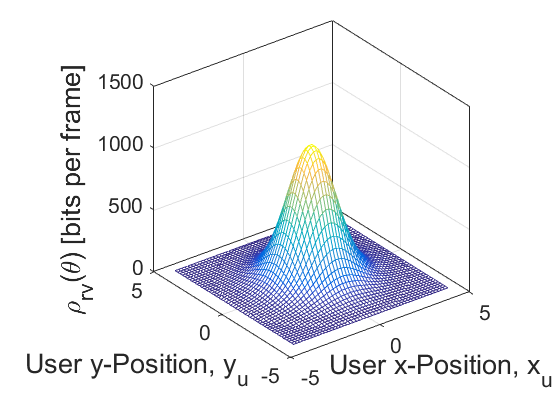}\label{max_avg_theta_60_path_1_theta_0_1_b}}\quad
	\subfigure[Hybrid-type II, $\theta = 0.1$]{\includegraphics[width=\figsize\textwidth]{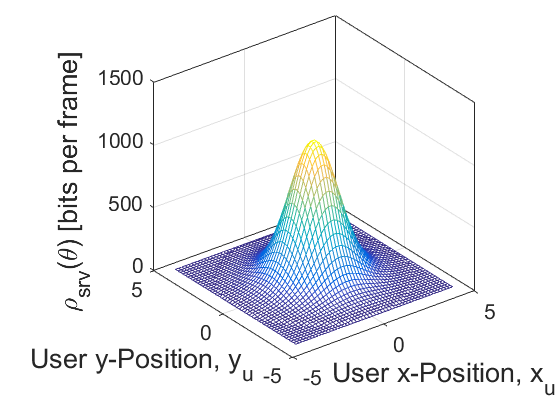}\label{max_avg_theta_60_path_2_theta_0_1_b}}
	\caption{Maximum average arrival rates for different selection strategies as a function of the receiver position in terms of $x_u$ and $y_u$ and for different values of $\theta$. Here, $P_{\text{avg}} = 24$ dBm, $\nu = 0.7$, $\alpha = 0.3$, $\beta = 0.7$ and $\phi_{1/2} = 60^{\circ}$. \{bpf : bits per frame\}.}\label{max_avg_theta_60}
	\vspace{-0.8cm}
\end{figure}
In \figurename~\ref{max_avg_theta_60}, we plot the maximum average arrival rate as a function of the $x_u$ and $y_u$ coordinates of the receiver location for different QoS constraints, where $x_{u}\in[-5,5]$ and $y_{u}\in[-5,5]$. As seen in \figurename~\ref{max_avg_theta_60_path_2_a} and \figurename~\ref{max_avg_theta_60_path_2_b}, the position of the receiver does not impact the performance levels in the RF link necessarily, i.e., the performance level stays almost constant when the receiver stays in the defined range. However, the performance levels in the other strategies are affected by the position of the receiver, and the maximum average data arrival rate increases as the receiver gets closer to the point $(x_v, y_v, z_v) = (0, 0, -d_v)$ because the constant transmission rate from the VLC access point to the receiver increases and the stochastic nature of the RF link is mitigated more with the increasing rate in the VLC link. As seen in \figurename~\ref{max_avg_theta_60_path_1_a} and 
\figurename~\ref{max_avg_theta_60_path_1_b}, the maximum average data arrival rate goes to zero in \emph{VLC-only Strategy} as the receiver goes out of the coverage area of the VLC access point. Similarly, as seen in \figurename~\ref{max_avg_theta_60_path_1_theta_0_1_a}, \figurename~\ref{max_avg_theta_60_path_2_theta_0_1_a}, \figurename~\ref{max_avg_theta_60_path_1_theta_0_1_b} and \figurename~\ref{max_avg_theta_60_path_2_theta_0_1_b}, the maximum average data arrival rate becomes equal to the one in \emph{RF-only Strategy} as the receiver goes out of the the coverage area of the VLC access point. Furthermore, \emph{Hybrid-type I Transmission Strategy} provides higher performance levels than \emph{RF-only Strategy} and \emph{VLC-only Strategy} do because the transmitter, employing \emph{Hybrid-type I Transmission Strategy}, sends the data over the RF link when the instantaneous transmission rate in the RF link is higher than the rate in the VLC link, and mitigates the lower transmission rates in the RF link by sending the data in the VLC link. Finally, \emph{Hybrid-type II Transmission Strategy} outperforms all the other strategies. However, the performance gap between \emph{Hybrid-type I Transmission Strategy} and \emph{Hybrid-type II Transmission Strategy} is not necessarily large. Hence, it is more advantageous to employ \emph{Hybrid-type I Transmission Strategy} in order to avoid the hardware complexity that follows the addition of multihoming capability in \emph{Hybrid-type II Transmission Strategy}.

\begin{figure}[t]
	\centering
	\subfigure[$(x_u,y_u,z_u) = (0,0,-d_v)$ i.e., cell center]{\includegraphics[width=\figsize\textwidth]{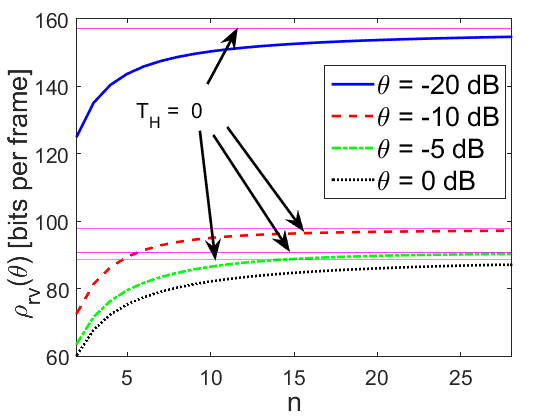}\label{fig:Rate_RF_VLC_handover_xu_0}}
	\subfigure[$(x_u,y_u,z_u) = (d_c,0,-d_v)$ i.e., cell edge]{\includegraphics[width=\figsize\textwidth]{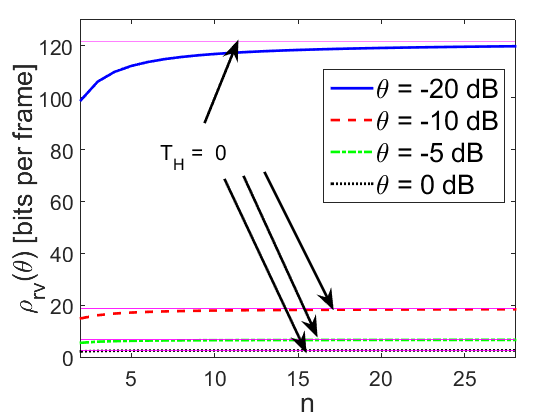}\label{fig:Rate_RF_VLC_handover_xu_rc}} 
	\caption{\small Maximum average arrival rate as a function of $n$ and for different values of the QoS exponent $\theta$ and user position in terms of $x_u$. Here, $\phi_{1/2} = 60^{\circ}$, $\alpha = 0.3$, $\beta = 0.7$, $P_{\text{avg}} = 24$ dBm, and $\nu = 1$.}\label{fig:Rate_RF_VLC_handover}
	\vspace{-0.8cm}
\end{figure}

\begin{figure}[t]
	\centering
	\subfigure[$P_{\text{avg}} = 24$ dBm]{\includegraphics[width=\figsize\textwidth]{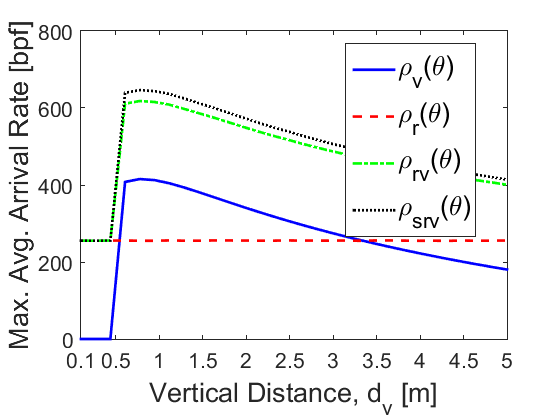}\label{fig:Rate_policies_dv_pavg_24}}
	\subfigure[$P_{\text{avg}} = 30$ dBm]{\includegraphics[width=\figsize\textwidth]{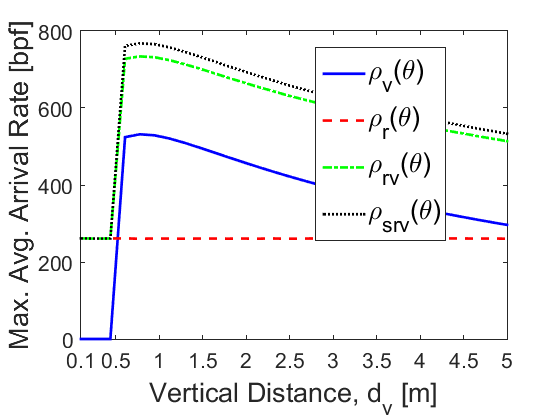}\label{fig:Rate_policies_dv_pavg_30}}
	\caption{\small Maximum average arrival rate of different transmission strategies as a function of the vertical distance and for different average power limit $P_{\text{avg}}$. Here, $\phi_{1/2}= 60^{\circ}$, $(x_u,y_u,z_u) = (0.8,0,-d_v)$, $\alpha = 0.3$, $\beta = 0.7$, $\theta = 0.1$, and $\nu = 1$. \{bpf : bits per frame\}.
}\label{fig:Rate_policies_dv}
\vspace{-0.8cm} 
\end{figure}

In \figurename~\ref{fig:Rate_RF_VLC_handover}, assuming that the handover process causes a transmission delay, where the handover process takes $T_{\text{H}}=\frac{1}{n}T$ seconds for $n \in \mathbb{N}$ and $n > 1$, we plot the maximum average data arrival rate in \emph{Hybrid-type I Transmission Strategy}, $\rho_{rv}(\theta)$, as a function of $n$ considering different user locations. Noting that smaller $n$ means a longer handover period, we observe that the transmission performance is highly affected by the handover process. With increasing $n$, the performance levels approach the values that are obtained when there is no handover delay. Moreover, the maximum average data arrival rates are higher in \figurename~\ref{fig:Rate_RF_VLC_handover_xu_0} than in \figurename~\ref{fig:Rate_RF_VLC_handover_xu_rc}, because the constant transmission rate in the VLC link is higher when the user is at the center. Subsequently in \figurename~\ref{fig:Rate_policies_dv}, we plot the maximum average data arrival rates as functions of the vertical distance between the VLC access point and the receiver. We set the position of the receiver to $(x_u,y_u,z_u) = (0.8,0,-d_v)$, i.e., the horizontal distance to the VLC cell center is $d_h = 0.8$ m. When $d_v \leq 0.6$ m, the performance level in the VLC link is zero because the cell area is very small and does not cover the point where the user stands, i.e., $d_c = d_v \tan(\theta_{1/2})< d_h = 0.8$ m, and $d_c$ is the cell radius. The performance levels in all strategies except \emph{RF-only Strategy} increase up to a value, and then decrease with increasing $d_v$. This is because the increase in the LED viewing angle is relatively less than the increase in $d_v$ at the beginning. Therefore, the user is effectively getting closer to the cell center and having more rate in the VLC link. In other words, the gain achieved by getting closer to the cell center is higher than the expected degradation due to the increasing cell radius. However, with $d_v$ increasing beyond a certain value, the user gets far away from the VLC access point, and hence, the radiated power spreads over more area, which eventually leads to decreased transmission rates in the VLC link. Therefore, the gain in the VLC link vanishes as the distance to the VLC access point becomes larger.

\begin{figure}[t]
	\centering
	\subfigure[RF-only Transmission, \mar{$\lambda$ = 21 kbpf}]{
	\includegraphics[width=\figsize\textwidth]{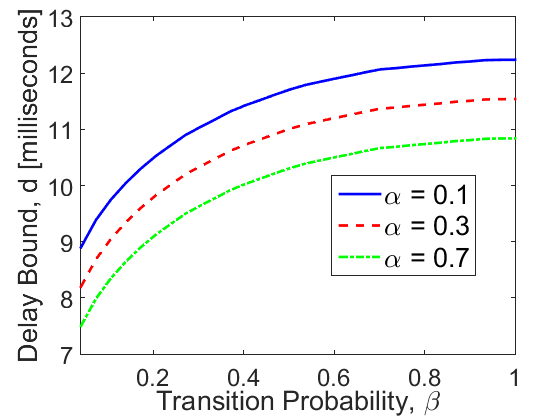}\label{delay_RF_beta_alpha}}
	\quad\quad
	\subfigure[VLC-only Transmission, \mar{$\lambda$ = 7 kbpf}]{
	\includegraphics[width=\figsize\textwidth]{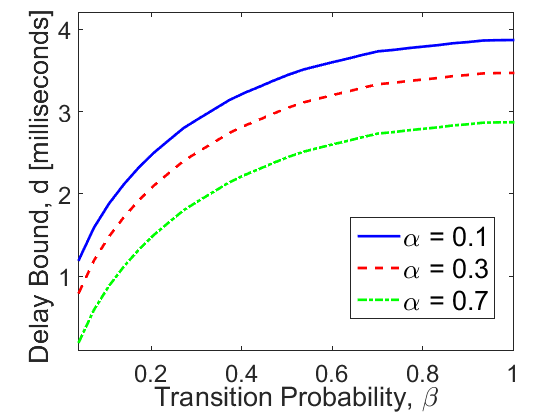}\label{delay_VLC_beta_alpha}}\\
	\subfigure[Hybrid-Type I Transmission, \mar{$\lambda$ = 24 kbpf}]{
	\includegraphics[width=\figsize\textwidth]{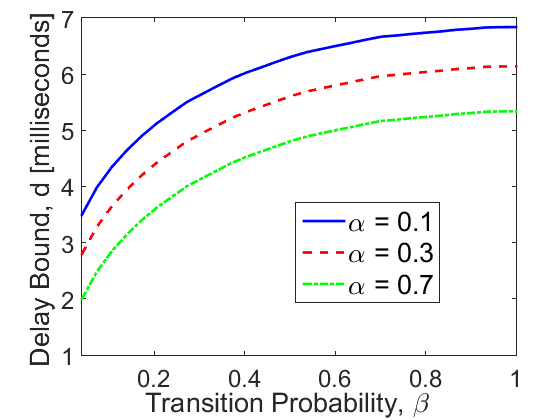}\label{delay_Hyb_I_beta_alpha}}
	\quad\quad
	\subfigure[Hybrid-Type II Transmission, \mar{$\lambda$ = 26 kbpf}]{
	\includegraphics[width=\figsize\textwidth]{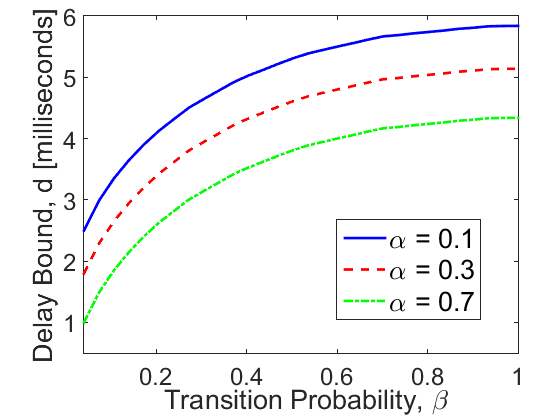}\label{delay_Hyb_II_beta_alpha}}
	\caption{Delay Bounds for different transmission strategies as a function of the transition probability $\beta$ and for different values of $\alpha$. Here, $d_0 = 10$ m, $d_1 = 3$ m, $P_{\text{avg}} = 30$ dBm, $\nu = 0.7$, and $\phi_{1/2} = 60^{\circ}$. \{bpf : bits per frame\}}\label{delay_beta_alpha}
	\vspace{-0.8cm}
\end{figure}
\vspace{-1cm}

\mar{\subsection{Non-asymptotic Delay Bounds}}
In the aforementioned results, we analyze the system performance in the steady-state. In the following, we provide results regarding the non-asymptotic bounds, i.e., the bounds on the buffering delay experienced by the data in the transmitter buffer. Particularly, we plot the delay bound as a function of the state transition probability from the OFF state to the ON state in the data arrival process, $\beta$, for different $\alpha$ values, where $\alpha$ is the state transition probability from the ON state to the OFF state in the data arrival process. We set the average data arrival rate, $\lambda p_{\text{ON}}$, to a value very close to the average data service (transmission) rate in the transmission channel. We note that the average data service rate in the transmission channel depends on the chosen transmission strategy. The delay bound is the highest in \emph{RF-only Strategy} as seen in \figurename~\ref{delay_RF_beta_alpha}, whereas it is the lowest in \emph{VLC-only Strategy} as seen in \figurename~\ref{delay_VLC_beta_alpha}. However, the arrival rate that \emph{RF-only Strategy} supports is higher than the rate that \emph{VLC-only Strategy} supports. More interestingly, the hybrid strategies can support higher arrival rates with less delay bounds, and \emph{Hybrid-type II Transmission Strategy} outperforms all the others. Herein, the system takes advantage of the occasional higher rates in the RF links, and mitigates the lower rates in the RF link by the constant transmission rate in the VLC link. Moreover, increasing $\beta$ and decreasing $\alpha$ cause the delay bound to increase.

\begin{figure}[t]
	\centering
	\subfigure[RF-only Transmission]{
	\includegraphics[width=\figsize\textwidth]{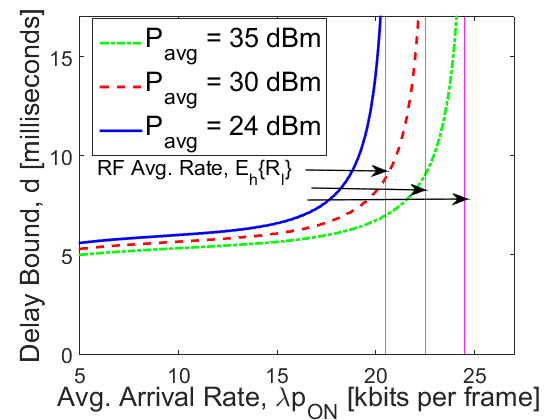}\label{Delay_RF_lambda}}
	\quad\quad
	\subfigure[VLC-only Transmission]{
	\includegraphics[width=\figsize\textwidth]{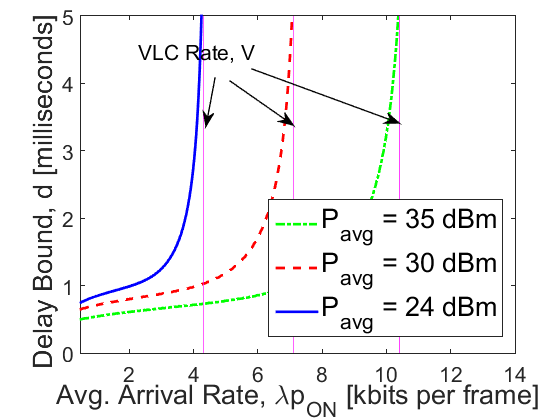}\label{Delay_VLC_lambda}}\\
	\subfigure[Hybrid-Type I transmission]{
	\includegraphics[width=\figsize\textwidth]{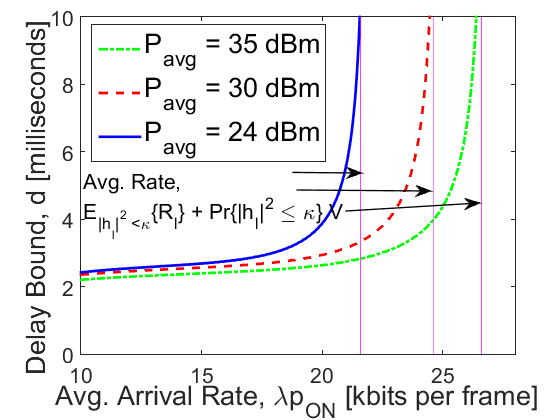}\label{Delay_Hyb_lambda}}
	\quad\quad
	\subfigure[Hybrid-Type II transmission]{
	\includegraphics[width=\figsize\textwidth]{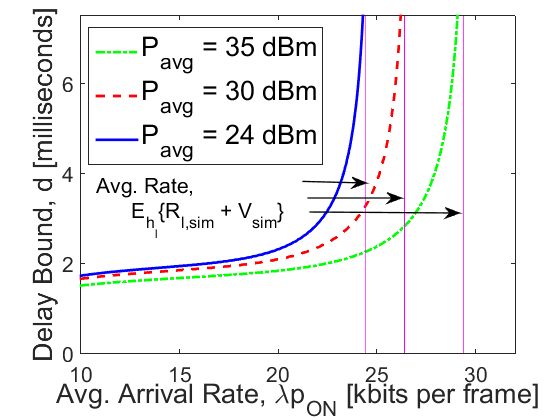}\label{delay_sim_lambda}}
	\caption{Delay Bounds for different transmission strategies as a function of the arrival rate $\lambda$ and for different values of $P_{\text{avg}}$. Here, $\alpha = 0.3$, $\beta = 0.7$, $d_0 = 10$ m, $d_1 = 3$ m, $\nu = 0.7$, and $\phi_{1/2} = 60^{\circ}$.}\label{delay_lambda}
	\vspace{-0.8cm}
\end{figure}
Finally, we explore the effects of the data arrival rate, $\lambda$, on the delay bound performance in \figurename~\ref{delay_lambda}. We set $\alpha = 0.3$ and $\beta = 0.7$, and consider different average power constraints, i.e., $P_{\text{avg}}=24$, $30$, and $35$ dBm. The delay bounds increase asymptotically as the average arrival rate approaches the average data service rates in the channels, because when the average data arrival rate is greater than the average data service rate in one channel, the system becomes unstable, and long buffering periods are expected. Moreover, as seen in \figurename~\ref{Delay_VLC_lambda}, the delay bounds are the minimum in \emph{VLC-only Strategy} but the ranges of the average data arrival rates are smaller than the other strategies, and as seen in \figurename~\ref{Delay_RF_lambda}, the delay bounds are the maximum in \emph{RF-only Strategy}. However, the hybrid strategies again outperform the others i.e., \mar{the hybrid strategies take advantage of the VLC link when the rate in the RF link goes down drastically, and utilize the RF link when the channel conditions are better again. While the VLC link provides stability and decreases the delay bounds, the RF link increases the range of average data arrival rate that can be supported.} Finally, in \figurename~\ref{delay_lambda} we see that increasing the average power can potentially improve the system performance in terms of the buffering delay.

\section{Conclusions}\label{sec:conc}
In this paper, we have analyzed the performance of a hybrid RF/VLC system when statistical QoS constraints are inflicted as limits on the buffer overflow and delay violation probabilities. We have provided a cross-layer analysis regarding physical and data-link layers by employing the maximum average arrival rate at the transmitter buffer and the non-asymptotic delay bounds as the main performance measures. We have proposed and analyzed three strategies in which RF and VLC links are utilized for data transmission. We have further formulated the performance levels achieved by each of the proposed strategies. Through numerical results, we have shown that RF technology can be beneficial when there are lower average power constraints and/or looser QoS requirements. Moreover, we have shown that utilizing the VLC technology for data transmission, either alone or in a hybrid transmission strategy, can potentially enhance the system performance in terms of delay performance. It lowers the buffering delay bounds, when compared to the RF technology. \mar{Particularly, when data arrival rates at the transmitter buffer is low, VLC links provide lower queueing delays than RF links do, but RF links support higher data arrival rates at the transmitter buffer.}

\appendix 
\vspace{-1.4cm}
\mar{\subsection{Derivation of $\rho_r(\theta)$ in (\ref{eq:rho_r})}\label{app:der_rho_r}
Recall that $\lambda$ is the data arrival rate at the buffer when the data source is in the ON state, and that $p_{\text{ON}}=\frac{\beta}{\alpha+\beta}$ is the steady-state probability of the ON state. Thus, the average data arrival rate is $\lambda p_{\text{ON}} = \frac{\beta}{\alpha+\beta}\lambda$. Moreover, we have $\Lambda_{a}(\theta^{\star}_{r})=-\Lambda_{r}(-\theta^{\star}_{r})$ in steady-state, i.e.,
\begin{equation}
	\begin{aligned}
		\log_{\text{e}}&\left\{\frac{1-\beta+(1-\alpha)e^{\theta^{\star}_{r}\lambda}+\sqrt{\left[1-\beta+(1-\alpha)e^{\theta^{\star}_{r}\lambda}\right]^2-4(1-\alpha-\beta)e^{\theta^{\star}_{r}\lambda}}}{2}\right\} = - \log_{\text{e}} \mathbb{E}_{h}\left\{e^{-\theta^{\star}_{r} R_{l}}\right\}.
	\end{aligned}
\end{equation}
Solving the aforementioned equation for $\lambda$ with any given $\theta>0$, we obtain
\begin{equation}
	\lambda = \frac{1}{\theta}\log_{\text{e}}\left\{\frac{1-(1-\beta)\mathbb{E}_{h}\left\{e^{-\theta R_{l}}\right\}}{(1-\alpha)\mathbb{E}_{h}\left\{e^{-\theta R_{l}}\right\}-(1-\alpha-\beta)\mathbb{E}^{2}_{h}\left\{e^{-\theta R_{l}}\right\}}\right\}.
\end{equation}
As a result, we formulate the maximum average data arrival rate as $\rho_{r}(\theta)=\frac{\beta}{\alpha+\beta}\lambda$ for any $\theta>0$. As for $\rho_{v}(\theta)$, we set $\Lambda_{a}(\theta^{\star}_{v})=-\Lambda_{v}(-\theta^{\star}_{v})$ and follow the same steps.}

\subsection{Proof of Proposition \ref{prob:prob_v} \label{app:prop_1}}
Based on the link selection process, the VLC link is selected only when $\rho_{v}(\theta) > \rho_{r}(\theta)$, where $\rho_{v}(\theta)$ is the maximum average arrival rate supported by the VLC link given in \eqref{eq:rho_v} and $\rho_{r}(\theta)$ is the maximum average arrival rate supported by the RF link given in \eqref{eq:rho_r}. Since logarithm is a monotonic increasing function, this condition is satisfied when we have
\begin{equation}\label{OF_Proof_1}
\frac{e^{2\theta V}-(1-\beta)e^{\theta V}}{(1-\alpha)e^{\theta V}-(1-\alpha-\beta)} > \frac{1-(1-\beta)\mathbb{E}_{h}\left\{e^{-\theta R_{l}}\right\}}{(1-\alpha)\mathbb{E}_{h}\left\{e^{-\theta R_{l}}\right\}-(1-\alpha-\beta)\mathbb{E}^{2}_{h}\left\{e^{-\theta R_{l}}\right\}},
\end{equation}
Now, let $\chi = \mathbb{E}_{h}\{e^{-\theta R_{l}}\}$ and $\mathcal{O} = e^{\theta V}$. Then, (\ref{OF_Proof_1}) can be expressed by the following quadratic inequality: 
\begin{equation}
\label{OF_Proof_3}
\mathcal{O}^2 - (1-\beta + (1-\alpha) \xi) \mathcal{O} + \xi(1-\alpha - \beta) > 0. 
\end{equation} 
where $\xi= \frac{1-(1-\beta)\chi}{(1-\alpha)\chi-(1-\alpha-\beta)\chi^2}$. \mar{Solving the above equation results in two solutions:
\begin{equation}
\mathcal{O}_1 = \frac{1-\beta+(1-\alpha)\xi - \sqrt{\left[1-\beta+(1-\alpha)\xi\right]^2-4(1-\alpha-\beta)\xi}}{2}
\end{equation}
and 
\begin{equation}
\mathcal{O}_2 =\frac{1-\beta+(1-\alpha)\xi + \sqrt{\left[1-\beta+(1-\alpha)\xi\right]^2-4(1-\alpha-\beta)\xi}}{2}
\end{equation}
where $\mathcal{O}_2 > \mathcal{O}_1$. $V$ has two ranges $0 < V < \log_{\text{e}} \{\mathcal{O}_1\}$ and $V \geq \log_{\text{e}} \{\mathcal{O}_2\}$. Setting $\mathcal{O} = 1$ in (\ref{OF_Proof_3}), we have $\beta (1-\xi)>0$. Note that $\xi > 1$ because $\rho_r(\theta) = \frac{\beta}{(\alpha + \beta)\theta} \log_{\text{e}} \{\xi\}>0$. Hence, we have $\beta (1-\xi) <0$, which implies that $\mathcal{O}_1 < 1$ and $\log_{\text{e}} \{\mathcal{O}_1\} < 0$. Therefore, we have only $V \geq \log_{\text{e}} \{\mathcal{O}_2\}$ as the solution region, which completes the proof.}
%\vspace{-0.5cm}

\bibliographystyle{IEEEtran}
\bibliography{references}

\end{document}